\documentclass[final,3p,times,review]{elsarticle} 

\usepackage{amssymb,amsmath,array} 

\usepackage{float} 

\usepackage[font=normal]{caption} 

\usepackage[flushleft]{threeparttable} 
\usepackage{multirow}
\usepackage{lscape}
\usepackage{pdflscape}
\usepackage{amsmath}
\usepackage{makecell}

\usepackage{subcaption}
\usepackage{graphicx}

\usepackage{nomencl}
\makenomenclature
\usepackage{etoolbox}
\renewcommand\nomgroup[1]{%
	\item[\bfseries
	\ifstrequal{#1}{A}{Symbols}{%
		\ifstrequal{#1}{B}{Subscripts}{%
  		\ifstrequal{#1}{C}{Superscripts}{%
			\ifstrequal{#1}{D}{Abbreviations}{}}}}%
	]}

\usepackage{soul} 

\usepackage{color} 

\usepackage{xcolor} 

\definecolor{mc1}{rgb}{0,0,0} 

\journal{Applied Energy} 

\usepackage{lineno} 

\begin{document} 



\begin{frontmatter}

\title{A novel control strategy to neutralize heat source within solid oxide electrolysis cell (SOEC) under variable solar power conditions}

\author[add1]{Zhaojian Liang} 
\author[add1]{Shanlin Chen} 
\author[add2]{Meng Ni} 
\author[add3]{Jingyi Wang\textsuperscript{**,}}
\ead{wangjingyi@hit.edu.cn}
\author[add1]{Mengying Li\textsuperscript{*,**,}}
\ead{mengying.li@polyu.edu.hk} 

\fntext[fn1]{\textsuperscript{*}Corresponding author.}
\fntext[fn2]{\textsuperscript{**}The two authors have the same contribution to this study.}

\address[add1]{Department of Mechanical Engineering \& Research Institute for Smart Energy, The Hong Kong Polytechnic University, Hong Kong SAR}
\address[add2]{Department of Building and Real Estate, Hong Kong Polytechnic University, Hung Hom, Kowloon, Hong Kong, China}
\address[add3]{School of Science, Harbin Institute of Technology, Shenzhen, 518055, China}

\begin{abstract} 
The integration of a solid oxide electrolysis cell (SOEC) with a photovoltaic (PV) system presents a viable method for storing variable solar energy through the production of green hydrogen. To ensure the SOEC's safety and longevity amidst dramatic fluctuations in solar power, control strategies are needed to limit the temperature gradients and rates of temperature change within the SOEC. Recognizing that the reactant supply influences the current, a novel control strategy is developed to modulate heat generation in the SOEC by adjusting the steam flow rate. The effectiveness of this strategy is assessed through numerical simulations conducted on a coupled PV-SOEC system using actual solar irradiance data, recorded at two-second intervals, to account for rapid changes in solar exposure. The results indicate that conventional control strategies, which increase airflow rates, are inadequate in effectively suppressing the rate of temperature variation in scenarios of drastic solar power changes. In contrast, our proposed strategy demonstrates successful management of the SOEC's heat generation, thereby reducing the temperature gradient and rate of variation within the SOEC to below 5 K cm$^{-1}$ and 1 K min$^{-1}$, respectively.
\end{abstract}

\begin{highlights} 

\item Modeled a coupled Photovoltaic-Solid Oxide Electrolysis Cell (PV-SOEC) system operating under real-world solar conditions.
\item Proposed a method to balance the heat source by modulating the steam supply rate.
\item Introduced control strategies of the heat source to minimize temperature variability.
\item Achieved a balance between thermal uniformity and electrolysis efficiency with minor endothermic operation.


\end{highlights}


\end{frontmatter} 

\section{Introduction} 
Renewable energy sources, notably solar and wind power, are augmenting their significance in today's energy structure \cite{WANG2023516}. It is expected that renewables will account for 80\% of new power capacity to 2030, with
solar PV alone contributing more than 50\%~\cite{iea2023world}. However, the integration of renewable energy still faces challenges. For instance, the intermittence and variability of power generation can introduce huge difficulties in the system operation and management (e.g., a mismatch between the generation and load). 
The uncertainty in power generation necessitate a time-spanning storage system, where green hydrogen emerges as a potential solution to these challenges, serving as a promising seasonal energy storage medium. Due to its long-term chemical stability, transportability, and carbon-free energy conversion \cite{HU2022118788}, hydrogen is increasingly viewed as a viable option for the storage of renewable energy. Consequently, power-to-hydrogen systems have been receiving significant attention in recent years \cite{MIN2022120145}.

The solid oxide electrolysis cell (SOEC) is a promising high-temperature electrolysis device for hydrogen production, owing to its high efficiency and operational flexibility. Notably, the SOEC can switch to a solid oxide fuel cell (SOFC) mode for power generation and adapt to various fuels \cite{XU2022115175,BEALE2021100902}. The integration of a photovoltaic (PV) system with SOECs offers an effective approach to store solar energy as hydrogen's chemical energy. In this way, PV panels harness solar energy to generate electrical power, which is subsequently utilized by the SOEC for water splitting and hydrogen production. However, the intermittent nature of solar power is a challenge to the thermal management of SOEC. During SOEC operation, temperature gradients can induce thermal stress \cite{PETIPAS20132957}, potentially leading to delamination and cracks in the electrolyte and electrodes \cite{ZENG2020115899}. Furthermore, the rapid temperature variation rate can result in thermal fatigue \cite{WANG2019255} and undermine durability of SOEC.

Numerous studies have investigated the dynamic thermal management of SOEC and SOFC in various scenarios, such as the operations under fluctuating wind \cite{WANG2019255} and solar power\cite{SUN2022115560}, the switch between fuel cell and electrolyzer \cite{XIAO2023120383,LIU2022115318}, and load changes\cite{UDAGAWA200846,ZHANG2022115544,ZHU2023121655}. Several parameters have been found to be effective for temperature control of SOEC and SOFC, including air flow rate  \cite{UDAGAWA200846,BOTTA2019636,AGUIAR2005136,LIU2022115318,ZHANG2022115544,LU2024117852}, steam/fuel flow rate \cite{LIU2023139000,CHEN2023117596,LU2024117852}, inlet temperature \cite{TSERONIS2012530,SRIKANTH2018473}, and inlet gas compositions \cite{SUN2022116310,ZENG2020115899,Tang2019,SUN2022102344}. 
For example, \citet{WANG2019255} modulated the air flow rate in response to wind power, successfully limiting the temperature gradient within the SOEC. However, they discovered that a significantly high air flow rate is necessary to maintain the temperature variation rate below the targeted 1\,K\,min$^{-1}$ under highly fluctuating power conditions. Such a high air flow rate would necessitate substantial compressor work. 
\citet{SUN2022116310} suggested that a control method focusing on the heat source is the key for constraining temperature fluctuations within SOECs. Therefore, they integrated thermochemical energy storage systems with SOEC to balance variations in the heat source. However, the temperature fluctuations in SOEC remain significant under severe power variations. Although existing studies have demonstrated some success in temperature control, further exploration into control strategies is crucial for the thermal stability of SOEC under highly fluctuating renewable power. 

In light of the challenges from severe power fluctuations, the concept of thermal neutrality could offer a solution to the thermal management issues. Theoretically, SOEC can operate at the thermoneutral voltage (TNV), which reduces the total heat source to zero, potentially leading to a more uniform temperature field, though not necessarily perfectly uniform. At TNV, the endothermic effect of water-splitting reactions counterbalances Joule heating and overpotential heat generation \cite{WANG2019255}, thereby simplifying thermal management \cite{YANG2021129260,Chen2017}. In real life scenarios, TNV is not a constant value but a variable that changes with operating conditions \cite{WANG2019255}. 
This observation implies the possibility of neutralizing heat source in SOEC by adjusting operating conditions. Such operational strategies would maintain the SOEC near the thermoneutral state under varying power supply, thereby improving the thermal uniformity and stability of SOEC. 
Despite this, there is a gap in literature concerning the operating strategy aimed at regulating the heat source. 

To fill the research gap, we introduce a strategy for controlling the heat source in SOEC under fluctuating solar power. This is achieved by modulating the steam supply rate according to real-time solar irradiance level.  The effectiveness of the proposed control strategy is evaluated using a three-dimensional transient numerical model of SOEC. Besides, we notice that the renewable power data utilized in previous studies are often recorded with low temporal resolution, such as 5 minutes \cite{SUN2022102344} or 15 minutes \cite{WANG2019255,XIAO2023120383}. The large time interval in solar data, however, overlooks the highly dynamic nature of renewable power sources. For instance, the solar irradiance can drop from 1000\,W\,m$^{-2}$ to 300\,W\,m$^{-2}$ within several seconds due to cloud attenuation. Thus, using solar data with low temporal resolutions could fail to capture these abrupt changes in irradiance, thereby potentially skewing the evaluation of PV-SOEC performance. To address this gap, we captured solar irradiance data at a significantly smaller time interval -- every 2 seconds. This high-resolution data was then used to assess the effectiveness of our proposed control strategy. 

In the following, Section\,\ref{Sec:methods} introduces the methodologies employed in this study, including the data acquisition and numerical models; Section\,\ref{Sec:results} provides comprehensive derivation and discussion on the proposed control strategy. Overall conclusion follows in Section~\ref{Sec:conclusion}.

\section{Methods} \label{Sec:methods}
\begin{figure}[h]
    \centering
    \includegraphics[width=\textwidth]{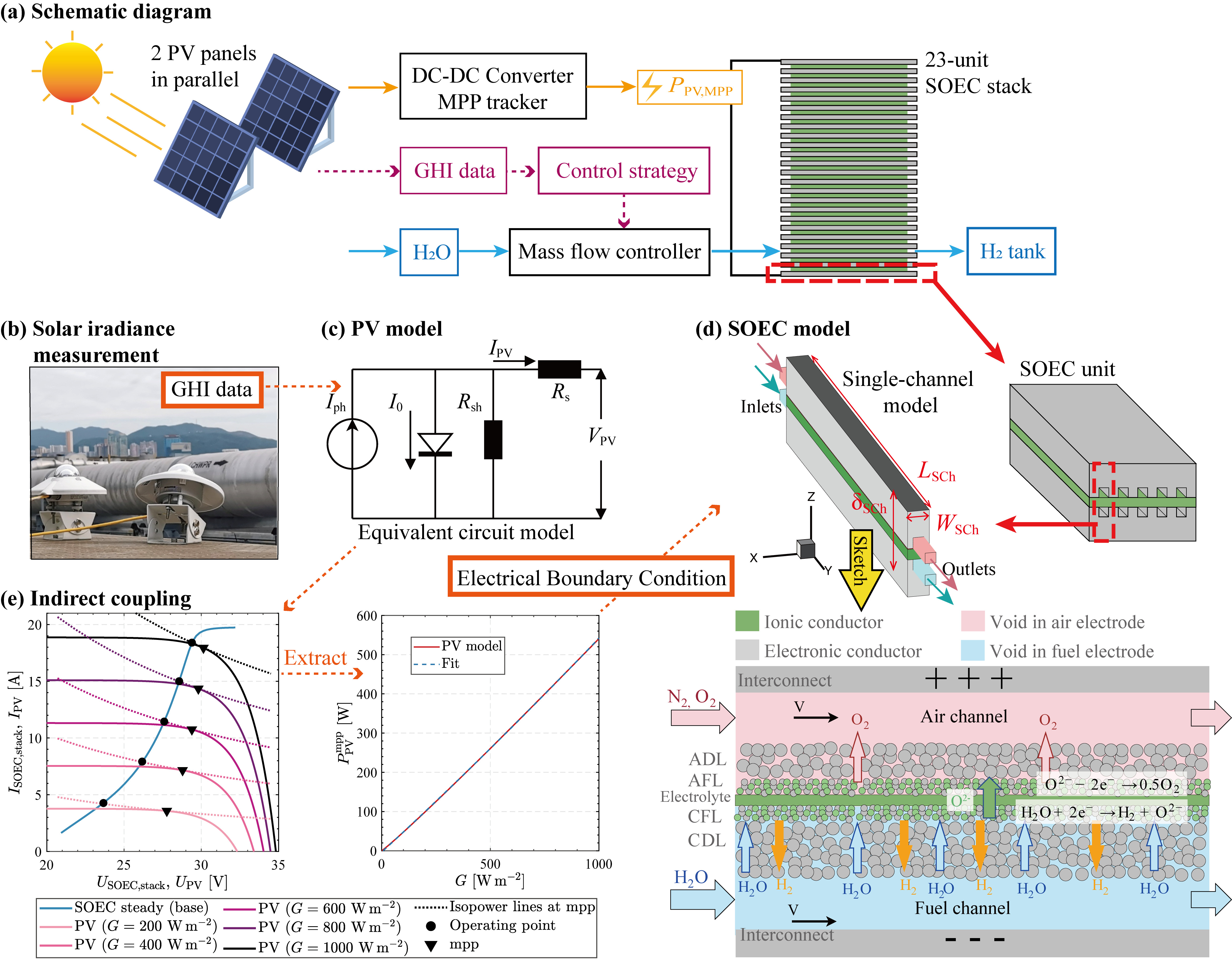}
    \caption{Overview of the methodology used in this study. \textbf{a} The schematic diagram of the indirectly coupled PV-SOEC system. \textbf{b} Measurement of solar irradiance on a rooftop solar monitoring station located on our campus. \textbf{c} Equivalent circuit model of the PV system \cite{LIAO2021119798}. \textbf{d} The geometry and working principle of SOEC. \textbf{e} The indirect coupling between PV and SOEC. The images of PV panels and hydrogen tank in \textbf{a} are sourced from \cite{hydrogenset}; partial images in \textbf{d} are modified from our previous work \cite{LIANG2023116759,liang2023characteristic}.}
    \label{fig:methodology}
\end{figure}

As shown in Fig.\,\ref{fig:methodology}a, this study aims to control the temperature of SOEC under fluctuating solar power supply. To numerically evaluate the effectiveness of our control strategy, we simulate an integrated PV-SOEC system as a case study. The PV panels convert the solar irradiance into electrical power, which is subsequently used in the SOEC for water splitting to produce hydrogen. Note that here we assume the PV panels are placed horizontally so it will utilize global horizontal irradiance (GHI) instead of plane-or-array irradiance. In our numerical model, the measured solar irradiance is converted into solar power data through a PV resource-to-power model. The solar power is then applied on our SOEC model for numerical simulation. The proposed control strategy is to dynamically adjust the steam flow rate of SOEC in response to solar power, ensuring the thermal safety of SOEC. 

\subsection{Solar irradiance measurement}
The GHI is measured using a pyranometer with a field of view of 180$^\circ$ (CMP10 produced by Kipp\&Zonen, ISO 9060 spectrally flat Class A), installed on the rooftop located on the campus of The Hong Kong Polytechnic University (22.3046$^\circ$N, 114.1796$^\circ$E). The GHI data is collected every 2 seconds to generate a high-temporal-resolution dataset. Additionally, to capture the cloud field of the sky, sky images are captured every minute using a 360$^\circ$ fisheye camera (FE9380-HV produced by VIVOTEK Corporation).

\subsection{PV model}
The PV model utilized in our study is based on the configurations reported in Refs.~\cite{LIAO2021119798,PVpanel}. 
As depicted in Fig.\,\ref{fig:methodology}a, the PV system under investigation comprises two monocrstalline PV panels, each with a nominal power rating of 280\,W. The number of PV panels is designed based on the operation range of the SOEC stack, which will be detailed in Section\,\ref{Sec:IndirectCouple}.
An equivalent diode circuit model \cite{LIAO2021119798}, shown in Fig.\,\ref{fig:methodology}c, is utilized to calculate the operating current and voltage of a PV panel. This model operates under the assumption that the series internal resistance $R_{\rm s}$ is nearly zero and the shunt resistance $R_{\rm sh}$ approaches infinity. The mathematical model for the two-panel system can be expressed as Eq.\,(\ref{Eq:PV_IV}).
\begin{equation}
I_{\rm{PV}}=n_{\rm p}^{\rm PV}\left\{ I_{\rm{ph}}- I_0\left[\exp \left(\frac{q U_{\rm{PV}}}{m k_{\rm{B}} T_{\rm{PV}} n_{\rm s}^{\rm PV}}\right)-1\right] \right\}
\label{Eq:PV_IV}
\end{equation}
where, $n_{\rm p}^{\rm PV}=2$ represents the number of PV panels connected in parallel, $n_{\rm s}^{\rm PV}=60$ denotes the number of solar cells connected in series within a single PV panel. $q = 1.60 \times 10^{-19}$\,C is the elementary charge. $m = 1$ is the ideal factor of diode. $k_B = 1.38 \times 10^{-23}$\,J\,K$^{-1}$ is the Boltzmann constant. The photo-generated current $I_{\rm{ph}}$ is calculated as,
\begin{equation}
I_{\rm{ph}}=\frac{G}{G_{\rm{PV,ref}}}\left[I_{\rm{ph,ref }}+\mu\left(T_{\rm{PV}}-T_{\rm{PV,ref}}\right)\right]
\label{Eq:I_ph}
\end{equation}
where, $G_{\rm PV,ref} = 1000$\,W\,m$^{-2}$ is the reference solar radiation. $T_{\rm PV,ref} = 298$\,K is the reference temperature. $I_{\rm ph,ref} = 9.43$\,A is the photo-generation current under the reference solar irradiance. $\mu = 0.00047$\,A\,K$^{-1}$ is the short-circuit current temperature coefficient. The diode reverse saturation current $I_0$ is calculated as,
\begin{equation}
I_0=I_{0, {\rm ref }}\left(\frac{T_{\rm{PV}}}{T_{\rm{PV,ref}}}\right)^3 \exp \left[\frac{q E_{\rm{g}}}{k_{\rm{B}}}\left(\frac{1}{T_{\rm{PV,ref}}}-\frac{1}{T_{\rm{PV}}}\right)\right]
\label{Eq:I0}
\end{equation}
where, $I_{\rm 0,ref} = 1.25\times 10^{-9}$\,A is the diode reverse saturation current in the reference state. The band-gap $E_{\rm{g}}$ of the PV panel is calculated as \cite{LIAO2021119798},
\begin{equation}
E_{\rm{g}}=1.17-\frac{4.73 \times 10^{-4} T_{\rm{PV}}^2}{T_{\rm{PV}}+636}
\label{Eq:E_g}
\end{equation}

Combining Eqs.\,(\ref{Eq:PV_IV}-\ref{Eq:E_g}), it is convenient to acquire the current-voltage curve of the PV system given a solar irradiance value $G$, as shown in Fig.\,\ref{fig:methodology}e. The maximum power point (MPP) tracker can be used to operate the PV system at its MPP.
When $n_{\rm p}^{\rm PV}$, $n_{\rm s}^{\rm PV}$ and $T_{\rm{PV}}$ of the PV system are held constant, the maximum power $P_{\rm PV}^{\rm MPP}$ can be determined from Fig.\,\ref{fig:methodology}e and the following equation,
\begin{equation}
P_{\rm PV}^{\rm MPP}=I_{\rm PV}^{\rm MPP}U_{\rm PV}^{\rm MPP}
\label{Eq:power}
\end{equation}
The relationship between the maximum power $P_{\rm PV}^{\rm MPP}$ and solar irradiance $G$ can be extracted from Fig.\,\ref{fig:methodology}e and easily implemented into our numerical model by a regression correlation,
\begin{equation}
P_{\rm PV}^{\rm MPP}=3.4737 \times 10^{-5} \cdot G^2 + 0.5106 \cdot G - 3.4792
\label{Eq:PG}
\end{equation}

\subsection{SOEC single-channel model}
Figure\,\ref{fig:methodology}d shows the geometry and working principle of SOEC. The electron conductors within SOEC include interconnects, Anode Diffusion Layer (ADL), Anode Functional Layer (AFL), Cathode Functional Layer (CFL), and Cathode Diffusion Layer (CDL). The ADL, AFL, CDL, and CFL are porous medium for gas transport. At high temperatures, oxygen ions can travel within AFL, CFL, and the solid oxide electrolyte. In functional layers where electrons, ions, and gases gather, H$_2$O is split into hydrogen and oxygen via electrochemical reactions. 

To simulate the performance of the SOEC stack, we have developed a 3-D single-channel Computational Fluid Dynamics (CFD) model, ensuring a balance between computational cost and accuracy~\cite{BAE2019112152,Su2022}. The 3-D CFD model, depicted in Fig.\,\ref{fig:methodology}d, is an established model developed and validated in our previous research \cite{LIANG2023116759}. This model was built employing Ansys Fluent R18.1, a commercial CFD software. 

The model solves a coupled set of conservation equations for momentum, species mass, energy, electronic charge, and ionic charge. These solutions yield transient results for velocity, species mass fractions, temperature, and electrical and ionic potentials within SOEC. 
Our model \cite{LIANG2023116759} incorporates the following theoretical principles:
\begin{itemize}
    \item Fick's Law is employed to model gas diffusion within fluid channels. The process accounts for variations in temperature and pressure. In porous media, Extended Fick's law is used to consider the influence of microstructure on mass diffusivity.
    \item Electrochemical reactions are modeled using the Butler-Volmer equation.
    \item Ideal Gas Law is utilized for the calculation of fluid densities. Properties such as viscosity, specific heat, and thermal conductivity, at ambient pressure, are sourced from the NIST database \cite{nist} and integrated into the model using piecewise linear interpolation.
    \item For the solid materials, properties such as ionic, electronic, and thermal conductivities are assumed to be constant, except for the temperature-dependent ionic conductivity of the electrolyte. The effective properties of porous media are computed by volumetric averaging of the fluid and solid properties.
\end{itemize}

In this work, the employed boundary conditions (BC) are presented in Table\,\ref{Tab:BC}. The electrical BC is modified to enable the single-channel SOEC model to represent the coupling between the PV system and the SOEC stack. The detailed coupling method is explained in the following section. 

\begin{table*}[]
\centering
\caption{Boundary conditions of the numerical model utilized in this work. Note that the operating pressure is $p_0=1\,{\rm atm}$.}
\label{Tab:BC}
\small
\begin{threeparttable}
\begin{tabular*}{\textwidth}{@{\extracolsep{\fill}}lllllll}
\hline
 &
  \begin{tabular}[c]{@{}l@{}}Inlet\tnote{\textit{a}} \end{tabular} &
  \begin{tabular}[c]{@{}l@{}}Outlet \end{tabular} &
  \begin{tabular}[c]{@{}l@{}}$x=0$ ,\\ $x/W_{\rm SCh} = 1$\end{tabular} &
  \begin{tabular}[c]{@{}l@{}}$z=0$ \end{tabular} &
  \begin{tabular}[c]{@{}l@{}}$z/\delta_{\rm SCh}=1$ \end{tabular} &
  \begin{tabular}[c]{@{}l@{}}Other \\ surfaces\end{tabular} \\ \hline
Momentum   & \makecell[l]{Base: $V_{\rm in}^{\rm air}=2$, $V_{\rm in}^{\rm fuel}=2$ \\ Strategy A: $V_{\rm in}^{\rm air}=10$, $V_{\rm in}^{\rm fuel}=2$ \\ Strategy B: $V_{\rm in}^{\rm air}=2$, $V_{\rm in}^{\rm fuel}=$ Eq.\,(\ref{Eq:strategyB}) \\ Strategy C: $V_{\rm in}^{\rm air}=2$, $V_{\rm in}^{\rm fuel}=$ Eq.\,(\ref{Eq:strategyC})} & $p_{\rm gauge}=0$ & Zero flux &     N.A.             &         N.A.           & $V = 0$     \\
Thermal    & $T_{\rm in}=1073$\,K      &    O.B.   & Zero flux & Zero flux          & Zero flux            & Zero flux \\
Species    & \makecell[l]{$X_{\rm in}^{\rm O_2}=0.2$, $X_{\rm in}^{\rm N_2}=0.8$, $X_{\rm in}^{\rm H_2O}=0.9$, $X_{\rm in}^{\rm H_2}=0.1$}                                 &   O.B.    & Zero flux &       N.A.      &    N.A.       & Zero flux \\
Electrical &   N.A.  &    N.A.   & Zero flux & $\phi_{\rm ele} = 0$ & $i_{\rm{SOEC,SCh}}$\tnote{\textit{b}} & Zero flux \\ \hline
\end{tabular*}
\begin{tablenotes}
	\footnotesize
 	\item[\textit{a}] The cell encompasses two inlets: the fuel channel inlet and the air channel inlet. The fuel inlet solely comprises H$_2$O and H$_2$, while the air inlet solely comprises O$_2$ and N$_2$. The control of the SOEC operation is achieved by manipulating the inlet velocity [m\,s$^{-1}$].
	\item[\textit{b}] The electric current density $i$ on the surface is a time-dependent function, which ensures the maximum power output of the PV system is equal to the power input of the SOEC stack under fluctuating solar conditions. The details are introduced in Section\,\ref{Sec:IndirectCouple}.
 	\item[$*$] This numerical study compares four different control strategies applied to $V_{\rm in}$, with all other boundary conditions held constant. O.B. means `open boundary'. N.A. means `not applicable'. 
\end{tablenotes}
\end{threeparttable}
\end{table*}

\subsection{Indirect coupling between PV and SOEC} \label{Sec:IndirectCouple}
The coupling between PV and electrolyzer can be classified into direct and indirect coupling \cite{MAS2022116213}. In the case of direct coupling, the electrolyzer is directly linked to the PV system, which implies that both systems operate at identical current and voltage. However, it poses a challenge to maintain their operating point proximate to the MPP of PV, thereby reducing the efficiency of PV and the coupled system. To track the MPP of PV, the indirectly coupled PV-electrolyzer is integrated with a MPP tracker and a DC/DC converter \cite{Lin2022}, as shown in Figs.\,\ref{fig:methodology}a and e. In this way, the PV can be operated at MPP under various conditions, and the power transmitted to the electrolyzer approximates the maximum power output of the PV. Therefore, the indirect coupling approach is selected in this work. 

The PV-SOEC system, as shown in Fig.\,\ref{fig:methodology}a, consists 2 PV panels connected in parallel ($n_{\rm p}^{\rm PV}=2$) and an SOEC stack composed of 23 SOEC units connected in series  ($n_{\rm s}^{\rm SOEC}=23$). As illustrated in Fig.\,\ref{fig:methodology}e, the numbers of PV panels and SOEC units are tuned to ensure that SOEC system approximates 90\% of its max current when the PV is operating under $G = 1000$\,W\,m$^{-2}$. 
The maximum power output of the PV system is configured to equal the power input of the SOEC system,
\begin{equation}
P_{\rm SOEC,stack}=P_{\rm PV}^{\rm MPP}
\label{Eq:P_equal}
\end{equation}
In Fig.\,\ref{fig:methodology}e, the steady-state operating point of the PV-SOEC system is depicted as the intersection between the SOEC current-voltage curve and the iso-power line of the PV system.  

In the SOEC stack, each individual SOEC unit consists of five parallel channels, with a total active area of 10\,cm$^2$ ($A_{\rm{act}}^{\rm{SOEC,unit}}=10$\,cm$^2$). That said, the SOEC single-channel model has an active area of 2\,cm$^2$ ($A_{\rm{act}}^{\rm{SOEC,SCh}}=2$\,cm$^2$). To apply the electrical conditions from the stack level to the single-channel model, the following equation is used, 
\begin{equation}
I_{\rm{SOEC,SCh}}=I_{\rm{SOEC,unit}} \cdot \frac{A_{\rm{act}}^{\rm{SOEC,SCh}}}{A_{\rm{act}}^{\rm{SOEC,unit}}}=\frac{P_{\rm SOEC,stack}}{U_{\rm{SOEC,stack}}} \cdot \frac{A_{\rm{act}}^{\rm{SOEC,SCh}}}{A_{\rm{act}}^{\rm{SOEC,unit}}} = \frac{P_{\rm SOEC,stack}}{n_{\rm s}^{\rm SOEC}U_{\rm{SOEC,unit}}} \cdot \frac{A_{\rm{act}}^{\rm{SOEC,SCh}}}{A_{\rm{act}}^{\rm{SOEC,unit}}}
\label{Eq:I_SOEC_SCh}
\end{equation}
Then, the current density applied on the SOEC single-channel model can be computed as,
\begin{equation}
i_{\rm{SOEC,SCh}}= \frac{I_{\rm{SOEC,SCh}}}{A_{\rm{act}}^{\rm{SOEC,SCh}}}
\label{Eq:i_SOEC_SCh}
\end{equation}

Combining Eqs.\,(\ref{Eq:PG}-\ref{Eq:i_SOEC_SCh}), the electrical BC ($i_{\rm{SOEC,SCh}}$) of the SOEC model listed in Table\,\ref{Tab:BC} is determined. This BC is incorporated in Ansys Fluent by User Defined Function (UDF). 

The electrical efficiency of the SOEC is calculated as ~\cite{Tembhurne2019,HU2022118788},
\begin{equation}
\text{Electrical efficiency} = \frac{\text{Power consumed by electrolysis}}{\text{Power input of SOEC}} = \frac{|\Delta G^0_{\rm f}|\cdot (I_{\rm{SOEC,unit}} \cdot n_{\rm s}^{\rm SOEC})/(nF)}{P_{\rm SOEC,stack}} = \frac{U^0_{\rm eq}}{U_{\rm{SOEC,unit}}}
\label{Eq:eta_elec}
\end{equation}
where, $\Delta G^0_{\rm f} = -237.14 \times 10^3 \,{\rm J\,mol^{-1}}$ is the standard Gibbs free energy change of water formation, $n=2$ is the number of participating electrons for hydrogen production, $F = 96485\,{\rm C\,mol^{-1}}$ is the Faraday constant, $U^0_{\rm eq}=1.229\,{\rm V}$ is the equilibrium potential at standard condition \cite{Tembhurne2019}. 

The hydrogen production rate is,
\begin{equation}
\dot{Q}_{\rm H2} = \frac{3600\cdot I_{\rm SOEC,stack}M_{\rm H2}}{nF\rho_{\rm H2}^{\rm 0}}
\label{Eq:H2Rate}
\end{equation}
where, $\dot{Q}_{\rm H2}$ is the hydrogen production rate with unit L\,h$^{-1}$, $M_{\rm H2}=0.002$\,kg\,mol$^{-1}$ is the molar mass of hydrogen, $\rho_{\rm H2}^{\rm 0}=89.88$\,kg\,L$^{-1}$ is the hydrogen density under the standard condition.

\subsection{Control strategies of SOEC}
\citet{WANG2019255} suggested that keeping the magnitudes of temperature gradient and temperature variation rate below 5\,K\,cm$^{-1}$ and 1\,K\,min$^{-1}$ respectively could reduce the failure probability of SOEC. In this study, the temperature gradient is defined as,
\begin{equation}
\text{Temperature gradient} = \frac{T_{\rm out}-T_{\rm in}}{L_{\rm SCh}}
\label{Eq:Def_TGrad}
\end{equation}
The temperature variation rate is defined as, 
\begin{equation}
\text{Temperature variation rate} = \frac{T_{\rm ave}^{n}-T_{\rm ave}^{n-1}}{\Delta t}
\label{Eq:Def_TRate}
\end{equation}
where, the $T_{\rm ave}^{n}$ denotes the average temperature of SOEC at the $n$-th step. The $\Delta t$ is the simulation time step.

Increasing the air flow rate, denoted as Strategy A in this work (Table\,\ref{Tab:BC}), is a popular temperature control strategy proposed in the literature~\cite{LIU2022115318}. In Strategy A, the air flow rate is increased by five times when compared with base case. This strategy serves as a benchmark representing conventional control approaches. Strategies B and C adaptively control the steam flow rate to stabilize the heat source in SOEC accordingly to the instantaneous solar power level. The objectives of Strategies B and C are to maintain SOEC at states of thermal neutrality and slight endothermicity (the temperature gradient of -3\,K\,cm$^{-1}$), respectively. For Strategies B and C, the correlations between the steam velocity in fuel channel and input power are numerically regressed as (see Section\,\ref{Sec:strategyDetails} for more details),

\begin{equation}
    \text{Strategy B: }V_{\rm in}^{\rm fuel} = 0.3873 \cdot P_{\rm SOEC,SCh}-0.0326
    \label{Eq:strategyB}
\end{equation}
\begin{equation}
    \text{Strategy C: }V_{\rm in}^{\rm fuel} = 0.0179 \cdot P_{\rm SOEC,SCh}^2+ 0.323 \cdot P_{\rm SOEC,SCh} + 0.0355
    \label{Eq:strategyC}
\end{equation}

\section{Result and discussion}\label{Sec:results}

\subsection{Guidance from the transient characteristics of PV-SOEC}\label{Sec:trasientCharac}
Before delving into our heat source-based control strategies, we conducted a simulation to explain the transient characteristics of the PV-SOEC system in response to a sudden change of solar irradiance. The transient characteristics will provide guidance for the control strategies. 

In an indirectly coupled PV-SOEC system, fluctuations in solar irradiance can affect the power supply to the SOEC. These changes to the electrical state can, in turn, alter the heat and mass fields within the SOEC. Here, numerical simulation is utilized to illustrate the SOEC responses to a rapid ramp-down in GHI, characterized as a linear decrease from 1000 to 200\,W\,m$^{-2}$ in 1\,ms, and hold constant at 200\,W\,m$^{-2}$. The simulation adopted an adaptive time-step size to accurately capture responses in SOEC \cite{LIANG2023116759}. The BCs of the SOEC is presented as the base case in Table\,\ref{Tab:BC}. The animation of electrical responses of the coupled system is provided in the supplementary video S1. Figure \,\ref{fig:G1000to200} depicts the electrical, gaseous, and thermal responses of the SOEC to the sudden irradiance change. While the electrical parameters (including current, voltage, and power) can adapt swiftly to the GHI ramp-down, the gaseous response is slower, persisting for approximately 0.1\,s. The slowest response is that of temperature, which takes thousands of seconds to reach a new steady state. These differential speeds of electrical, gaseous, and thermal responses were previously discussed in our earlier studies \cite{LIANG2023116759,liang2023characteristic}. In light of the different response times, the fast-response factors such as current and gas flow rates can be employed to control the slow-response factors such as temperature. To adapt to the highly fluctuating solar power, effective operating strategies for temperature control should be based on the fast-response factors such as current and gas flow rates.

\begin{figure}[h]
    \centering
    \includegraphics[width=0.9\textwidth]{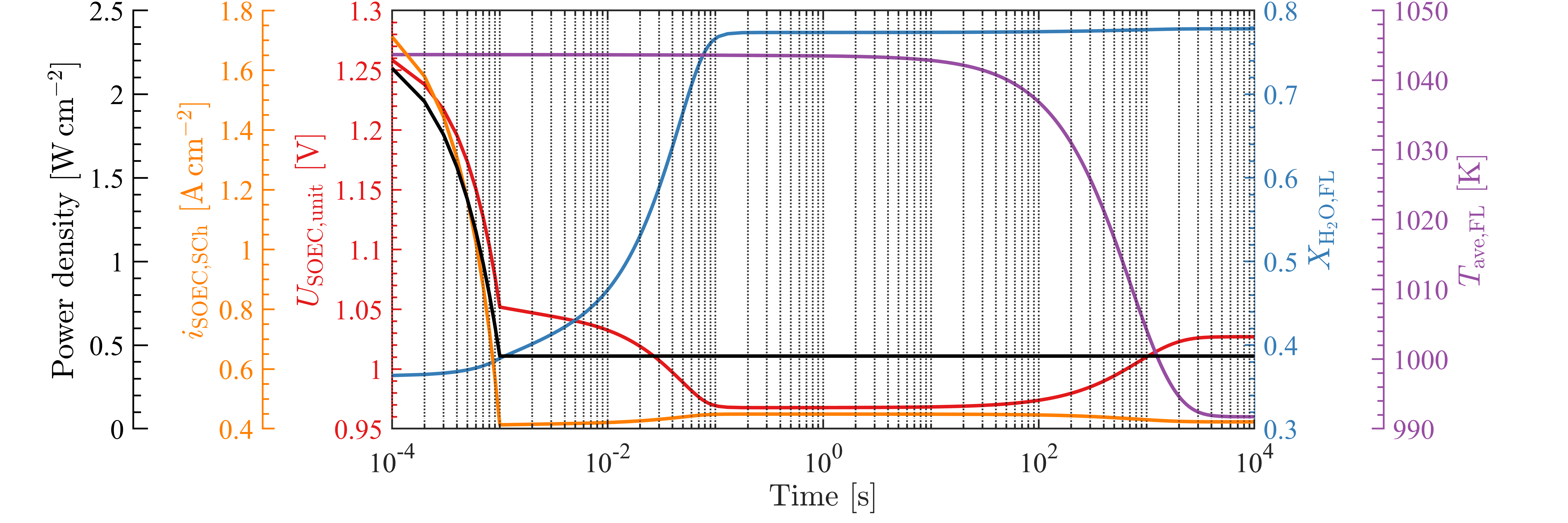}
    \caption{Responses of the single-channel SOEC model to a rapid ramp-down in GHI from 1000 to 200\,W\,m$^{-2}$ for a coupled PV-SOEC system. The power density, current density $i_{\rm SOEC,SCh}$, and voltage $U_{\rm SOEC,unit}$ variations with respect to time are illustrated. The time-varying average mole fraction of H$_2$O $X_{\rm H_2O,FL}$ and average temperature $T_{\rm ave,FL}$ in the functional layers are also presented.  }
    \label{fig:G1000to200}
\end{figure}

Figure\,\ref{fig:G1000to200} also shows a significant temperature drop of around 50\,K following the decrease of GHI. Figure\,\ref{fig:CFDResults}b-c compare the steady-state temperature distributions of the initial and final states of Fig.\,\ref{fig:G1000to200}. The temperature drop originates from the decrease of total heat source within the SOEC. Additionally, the negative total heat source induces temperature gradients within the SOEC. In order to reduce the failure probability of SOEC, it is vital to examine control strategies that can effectively manage the temperature gradient and variation of the SOEC, particularly when it's coupled with fluctuating power sources.

\subsection{Limitations of conventional control strategies based on air flow rate} 
As discussed in Section\,\ref{Sec:trasientCharac}, gas flow rate is a factor which has sufficiently fast response dynamics for the temperature control of SOEC. The strategy of controlling air flow rate has been widely adopted in literature \cite{UDAGAWA200846,BOTTA2019636,AGUIAR2005136,LIU2022115318,ZHANG2022115544}. However, this control strategy shows limited performance for the temperature control of SOEC under severe power fluctuations \cite{WANG2019255}. In this section, we utilize steady simulations to analyze the limitations of controlling air flow rate and point out the importance of controlling heat source. 

\begin{figure}
    \centering
    \includegraphics[width=0.9\textwidth]{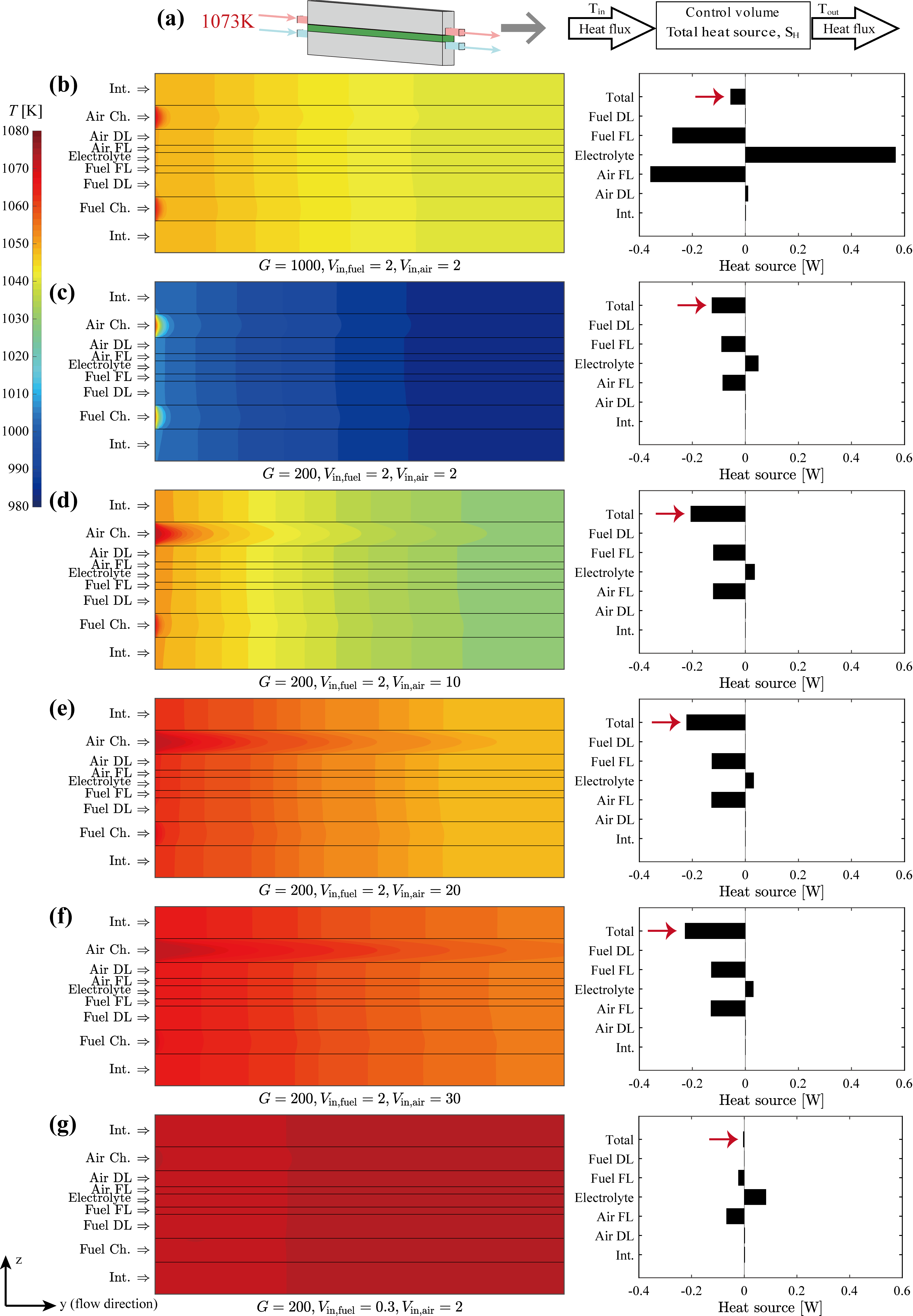}
    \caption{Heat transfer within SOEC.  \textbf{a} Energy conservation of the control volume. \textbf{b-g} Distributions of temperature (left panel) and heat source (right panel) under steady-state with different  solar irradiances ($G$ [W\,m$^{-2}$]) and inlet velocities of air and fuel channels  ($V_{\rm in, air}$, $V_{\rm in, fuel}$ [m\,s$^{-1}$]). The length scales of SOEC have been adjusted in this figure for better visualization.}
    \label{fig:CFDResults}
\end{figure}

Figure\,\ref{fig:CFDResults}c-f illustrate that, as the air velocity increases from 2 to 30\,m\,s$^{-1}$, the values of the total heat sources show minor difference, while the overall temperature of the SOEC significantly rises and the temperature gradient reduces. These observations, on one hand, suggest that augmenting the air flow rate is an effective strategy for reducing the temperature gradient. On the other hand, the thermal effect of increasing the air flow rate is similar to the `dilution' of the endothermic effect, without drastically altering the heat source within the SOEC. However, when the magnitude of heat source is large, a considerably high air flow rate is needed to dilute the heat source in order to ensure an acceptable temperature gradient \cite{WANG2019255}. This, in turn, necessitates substantial compressor work and rigorous gas-tight sealing of the SOEC. 

To further understand temperature control strategies, the SOEC can be considered as a control volume, as shown in Fig.\,\ref{fig:CFDResults}a. The steam and air flows represent the heat flux across the boundaries of the control volume. Within the control volume, electrochemical reactions and Joule heating function as the heat source $S_{\rm H}$. Upon reaching a thermal steady state within the control volume, the heat source is equal to the enthalpy difference between the heat fluxes, 
\begin{equation}
0=\dot{\mathcal{H}}_0-\dot{\mathcal{H}}+S_{\rm H} = \bar{\rho} \bar{c_p} \dot{\mathcal{V}}_{\rm ave}(T_{\rm in}-T_{\rm out})+S_{\rm H}
\label{Eq:EnergyConser}
\end{equation}
where $\dot{\mathcal{H}}_0$ and $\dot{\mathcal{H}}$ denote the enthalpy flow rates at the inlet and outlet, respectively. $\bar{\rho}$, $\bar{c_p}$, and $\dot{\mathcal{V}}_{\rm ave}$ denote the average density, specific heat capacity, and volumetric flow rate of the gases through the SOEC. The temperature difference between inlet and outlet can be further expressed as,
\begin{equation}
\Delta T_{\rm in,out} = T_{\rm out}-T_{\rm in} = \frac{S_{\rm H}}{\bar{\rho} \bar{c_p} \dot{\mathcal{V}}_{\rm ave}}
\label{Eq:DeltaT}
\end{equation}
Equation\,(\ref{Eq:DeltaT}) indicates the two general control strategies to reduce temperature difference $\Delta T_{\rm in,out}$:  one is to increase $\dot{\mathcal{V}}_{\rm ave}$, and the other is to reduce $S_{\rm H}$. The conventional control strategy based on air flow rate is equivalent to the adjustment of $\dot{\mathcal{V}}_{\rm ave}$. When $S_{\rm H}$ is high, adjusting $\dot{\mathcal{V}}_{\rm ave}$ shows reduced performance for constraining the $\Delta T_{\rm in,out}$ within SOEC due to the low specific heat capacity $c_p$ of air \cite{WANG2019255}. To effectively constrain $\Delta T_{\rm in,out}$, it is important to develop a control strategy that can control the heat source $S_{\rm H}$ in SOEC.

\subsection{Derivation of heat source-based control strategies}\label{Sec:heatsourceControl}
\subsubsection{Determination of key variables for controlling heat source}
\begin{figure}
    \centering
    \includegraphics[width=0.3\textwidth]{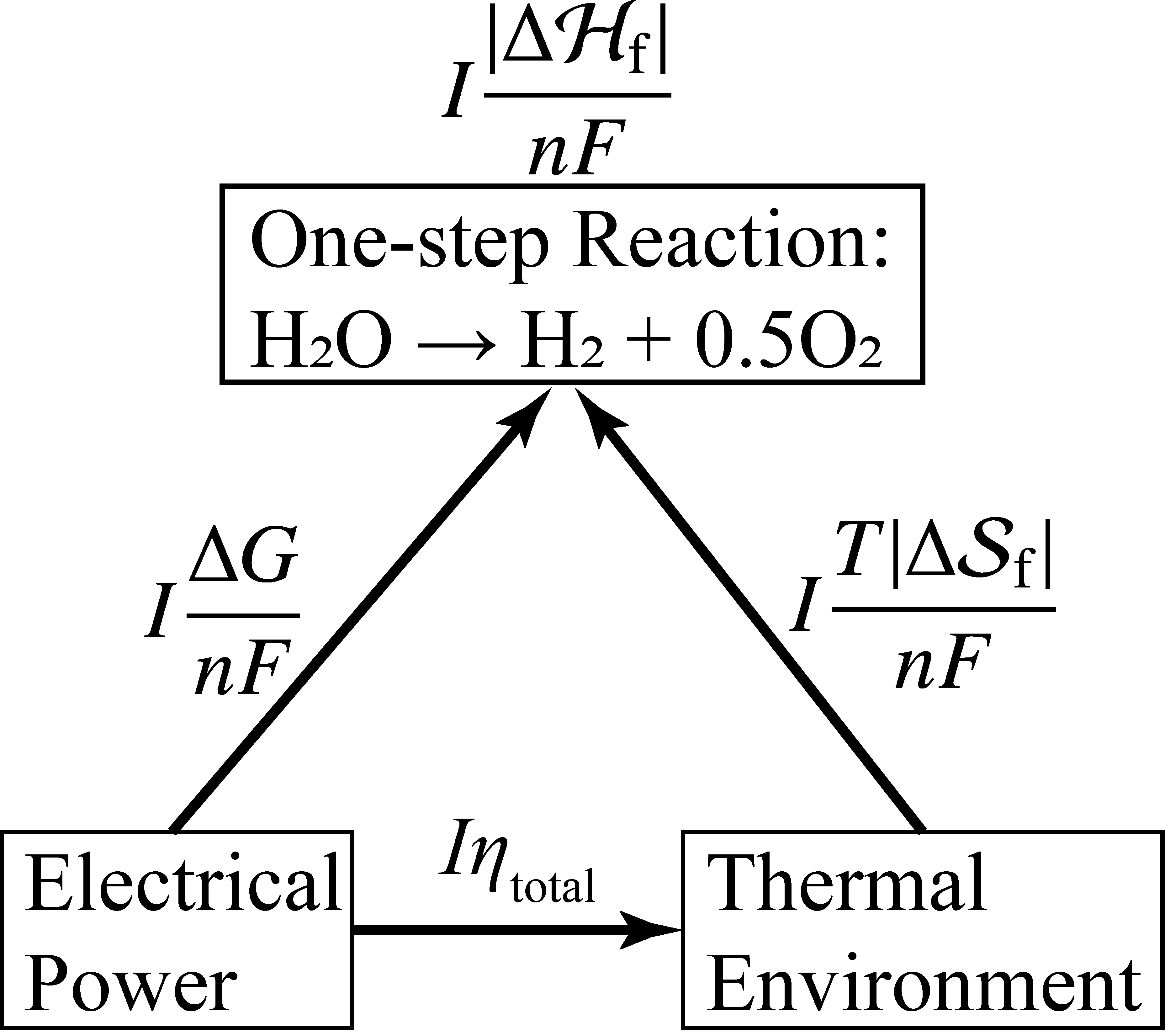}
    \caption{Energy balance in SOEC, which is simplified to one-step reaction. }
    \label{fig:EnergyBalance}
\end{figure}

In SOECs, the heat source is coupled with electrical power and electrochemical reactions. An in-depth understanding of the energy balance can facilitate the identification of key variables for controlling the heat source. As shown in Fig.\,\ref{fig:EnergyBalance}, the SOEC process is idealized as a one-step water-splitting reaction. The amount of energy required for the water-splitting reaction is $I |\Delta\mathcal{H}_{\rm f}|/nF$. Here, $\Delta\mathcal{H}_{\rm f}$ is the enthalpy of formation for water \cite{bard2022},
\begin{equation}
\Delta\mathcal{H}_{\rm f} = \Delta G + \Delta\mathcal{S}_{\rm f}
\label{Eq:DeltaH}
\end{equation}
where, $\Delta G$ is the change of Gibbs free energy, which is the minimum electrical energy required for electrolysis. $\Delta\mathcal{S}_{\rm f}$ is the entropy of formation of water. 

The water-splitting reaction acquires energy from electrical power source ($I \Delta G/nF$) and thermal environment ($I T |\Delta \mathcal{S}_{\rm f}|/nF$). Besides, the excess electrical power ($I\eta_{\rm total}$) input to SOEC is dissipated as heat, where $\eta_{\rm total}$ signifies the total overpotential applied to the SOEC. By examining the thermal energy balance, the total heat source $S_{\rm H}$ of SOEC can be determined,  
\begin{equation}
S_{\rm H} = I \eta_{\rm total} - I \frac{T |\Delta \mathcal{S}_{\rm f}|}{nF}
\label{Eq:HeatSource}
\end{equation}
The sign and magnitude of $S_{\rm H}$ determine the thermal behavior of the SOEC. From Eq.\,(\ref{Eq:HeatSource}) and Fig.\,\ref{fig:EnergyBalance}, we identify four variables that can be adjusted to control the magnitude of the heat source:
\begin{itemize}
    \item Reaction pathway. SOEC can accommodate different fuels \cite{XU2022115175}. By supplying different reactants, such as steam and carbon dioxide mixture, the $\Delta \mathcal{H}$ and $\Delta \mathcal{S}$ of reaction within the SOEC can be regulated \cite{LIU2023139000}. 
    \item Temperature \cite{SUN2022102344,SRIKANTH2018473}. Temperature has wide-ranging effects on factors such as the Ohmic resistance of the solid electrolyte, current, voltage, and enthalpy change. However, as discussed in Section\,\ref{Sec:trasientCharac}, the temperature responds more slowly than gas and electricity \cite{LIANG2023116759,TSERONIS2012530}. Thus, it might be challenging to regulate the heat source by only manipulating the temperature field, especially under conditions of highly fluctuating power.
    \item Current $I$. The change of current will vary the overpotential $\eta_{\rm total}$. The net current flowing through the system can be controlled by adjusting the series and parallel connections of SOEC units during operation \cite{PAUL2008490,MAROUFMASHAT201418743}. This, in turn, allows the regulation of the heat source.
    \item Inlet flow rate of the reactant. The maximum achievable current is limited by the supply of reactants in SOEC. Thus, we can control the heat source by adjusting the flow rate of reactant. The limiting current ($I_{\rm lim}$) is presented as,
    \begin{equation}
        I\leq I_{\rm lim} = \frac{n F p_0 X_{\rm in}^{\rm H_2O} A_{\rm ch}V_{\rm in}^{\rm fuel}} {\mathcal{R} T_{\rm in}}
        \label{Eq:limCurrent}
    \end{equation}
    where, $A_{\rm ch}=10^{-6}$\,m$^2$ is the cross area of fluid channel, $\mathcal{R}=8.314$\,J\,mol$^{-1}$\,K$^{-1}$ is the ideal gas constant. 
\end{itemize}
Considering the applicability of control strategies and the fast-response dynamics of gases (Section\,\ref{Sec:trasientCharac}), we propose to only adjust the inlet steam velocity at fuel channel ($V_{\rm in}^{\rm fuel}$) to continuously neutralize the heat source within the SOEC under varying power supply. 

\subsubsection{Theoretical feasibility of thermoneutral control strategy}
Theoretically, by adjusting $V_{\rm in}^{\rm fuel}$, the SOEC can be maintained at thermal neutrality for highly thermal uniformity. Based on Eq.\,(\ref{Eq:HeatSource}), SOEC achieves thermal neutrality ($S_{\rm H}=0$) when  
\begin{equation}
\eta_{\rm total} = \frac{T |\Delta\mathcal{S}_{\rm f}|}{nF}
\label{Eq:CancelOut}
\end{equation}
Then, based on Fig.\,\ref{fig:EnergyBalance}, the theoretical electrical power required to maintain SOEC at thermal neutrality can be expressed as, 
\begin{equation}
P_{\rm SOEC}^{\rm theo} = I \frac{\Delta G}{nF} + I \eta_{\rm total} = I \frac{\Delta G}{nF} + I \frac{T |\Delta \mathcal{S}_{\rm f}|}{nF} = I \frac{|\Delta\mathcal{H}_{\rm f}|}{nF}= IU_{\rm TN}^{\rm theo}
\label{Eq:EnergyCons_ideal}
\end{equation}
Consequently, the theoretical thermoneutral voltage $U_{\rm TN}^{\rm theo}$ for the water-splitting reaction \cite{godula2015hydrogen} is derived as follows,
\begin{equation}
U_{\rm TN}^{\rm theo} = \frac{|\Delta \mathcal{H}_{\rm f}|}{nF}
\label{Eq:U_TN}
\end{equation}
According to the NIST database \cite{nist}, at the ambient pressure and the temperature of 1073\,K, the values of $\Delta \mathcal{H}_{\rm f}$ and $U_{\rm TN}^{\rm theo}$ for water are -246.4\,kJ\,mol$^{-1}$ and 1.28\,V, respectively. 

To theoretically derive the thermoneutral control strategy, a high conversion rate of reactant H$_2$O is assumed at the thermal neutrality of SOEC and hence the current approximates the limiting current, 
\begin{equation}
    I \approx I_{\rm lim}
    \label{Eq:Iapprox}
\end{equation}
Combining Eqs.\,(\ref{Eq:limCurrent}-\ref{Eq:Iapprox}), the thermoneutral control strategy based on $V_{\rm in}^{\rm fuel}$ is theoretically derived,
\begin{equation}
    V_{\rm in,TN}^{\rm fuel} \approx \frac{\mathcal{R} T_{\rm in}}{n F p_0 X_{\rm in}^{\rm H_2O} A_{\rm ch} U_{\rm TN}^{\rm theo}} P_{\rm SOEC}^{\rm theo}
    \label{Eq:TNcontrolV}
\end{equation}
Given the values of $T_{\rm in}$, $X_{\rm in}^{\rm H_2O}$, and $p_0$ in Table\,\ref{Tab:BC}, Eq.\,(\ref{Eq:TNcontrolV}) can be written as a linear formula,
\begin{equation}
    V_{\rm in,TN}^{\rm fuel} \approx 0.3961 \cdot P_{\rm SOEC}^{\rm theo}
    \label{Eq:TNcontrolV_num}
\end{equation}
\subsubsection{Acquisition of heat source-based control strategies} \label{Sec:strategyDetails}
\begin{figure}
    \centering
    \begin{subfigure}{0.49\textwidth}
        \centering
        \includegraphics[width=\textwidth]{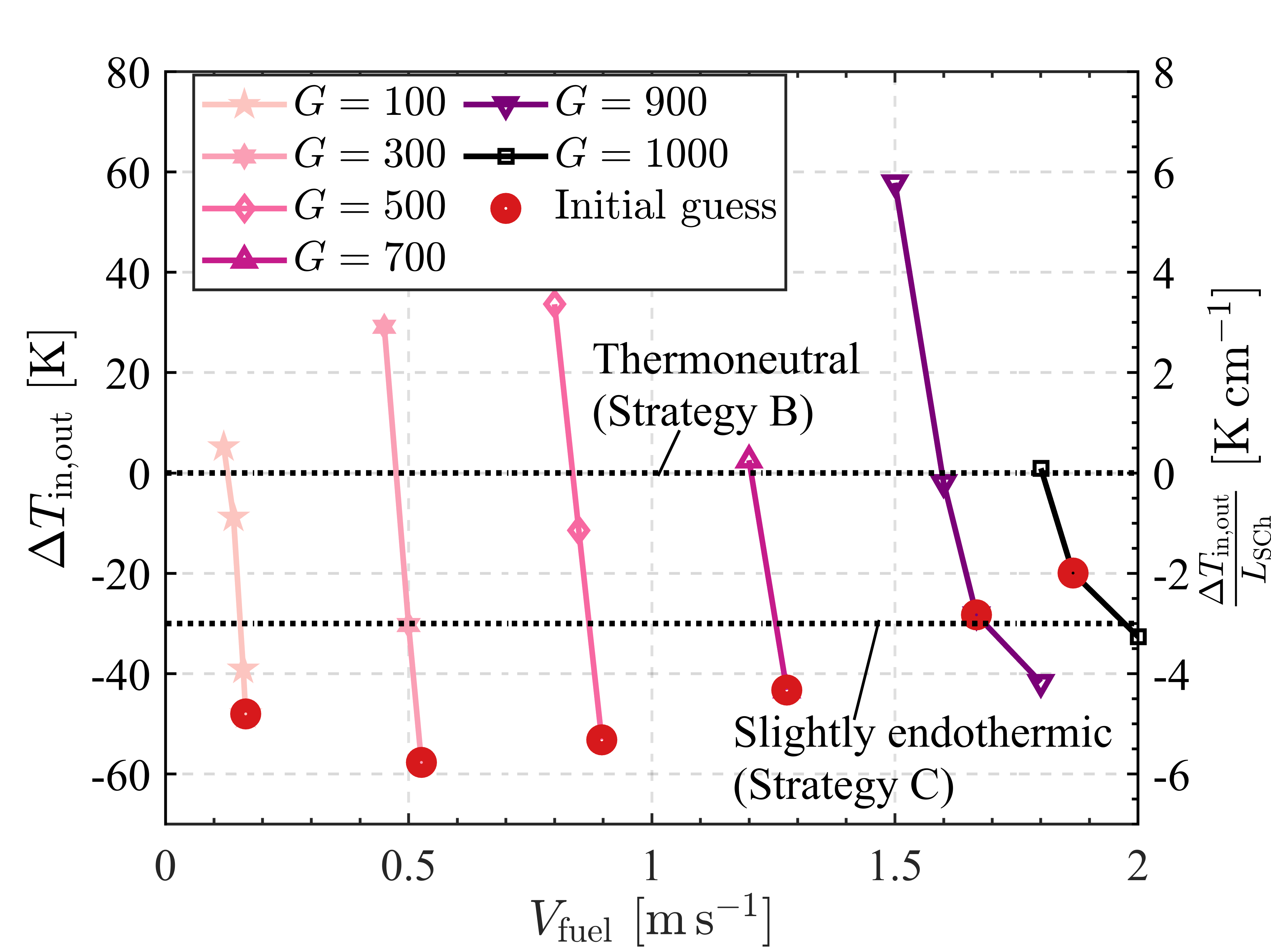}
        \caption{}
        \label{fig:deltaTV}
    \end{subfigure}
    \begin{subfigure}{0.49\textwidth}
        \centering
        \includegraphics[width=\textwidth]{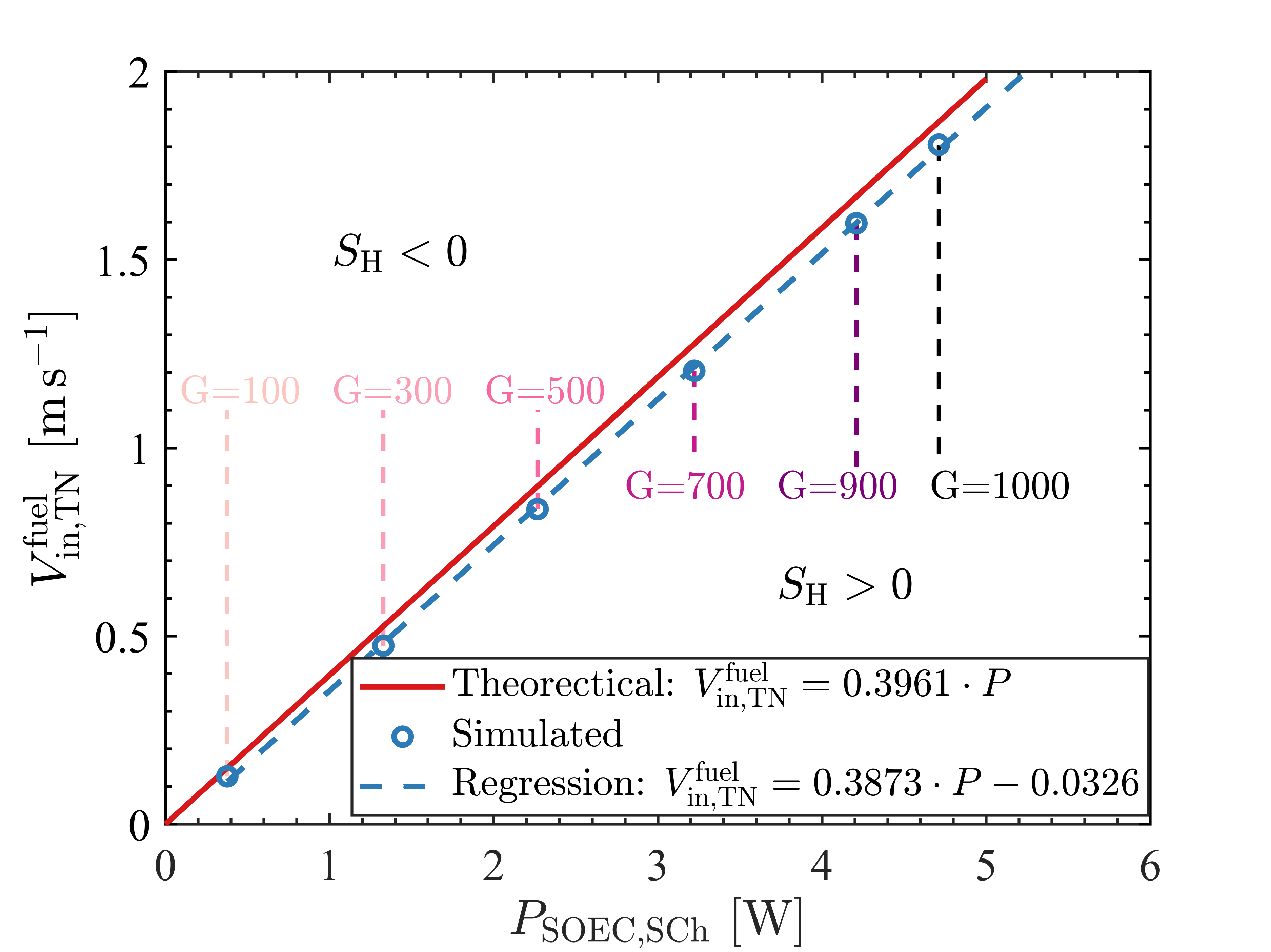}
        \caption{}
        \label{fig:VP}
    \end{subfigure}
    \caption{Steam velocity of fuel channel to achieve thermal neutrality in SOEC under different GHI ($G$[W\,m$^{-2}$]). (a) Temperature difference with respect to $V_{\rm in}^{\rm fuel}$ and $G$. The intersects between the color lines and the two dotted lines are the $V_{\rm in}^{\rm fuel}$ that could maintain the SOEC at the thermoneutral or slight endothermic state. (b) The thermoneutral $V_{\rm in}^{\rm fuel}$ with respect to power input of SOEC. }
    \label{fig:ControlV}
\end{figure}

Equation\,(\ref{Eq:TNcontrolV}) determines the theoretical feasibility of neutralizing the heat source within SOEC by adjusting $V_{\rm in}^{\rm fuel}$ under different power inputs. However, the theoretical derivation is based on assumptions such as Eq.\,(\ref{Eq:Iapprox}). Thus, $V_{\rm in,TN}^{\rm fuel}$ calculated from Eq.\,(\ref{Eq:TNcontrolV_num}) cannot exactly lead to the thermal neutrality of SOEC in reality, but gives the initial guess of the actual $V_{\rm in,TN}^{\rm fuel}$. Based on the initial guess, the $V_{\rm in}^{\rm fuel}$ to attain thermal neutrality ($\Delta T_{\rm in,out}=0,S_{\rm H}=0$) for different GHI levels are inferred by conducting a number of steady-state numerical simulations of the PV-SOEC system, as shown in Fig.\,\ref{fig:deltaTV}. Subsequently, the relationship between the thermoneutral flow rate $V_{\rm in,TN}^{\rm fuel}$ and input power $P_{\rm SOEC,SCh}$ is linearly regressed, as shown in Fig.\,\ref{fig:VP} and presented in Eq.\,(\ref{Eq:strategyB}). The thermoneutral control strategy, Eq.\,(\ref{Eq:strategyB}), is adopted as the Strategy B in this study. Furthermore, Fig.\,\ref{fig:VP} shows that the theoretically derived Eqs.\,(\ref{Eq:TNcontrolV}-\ref{Eq:TNcontrolV_num}) can not only provide a good guess of $V_{\rm in,TN}^{\rm fuel}$ but also serve as an indicator for the thermal states of SOEC. Given a value of $P_{\rm SOEC,SCh}$, SOEC is endothermic at the steam velocity greater than $V_{\rm in,TN}^{\rm fuel}$ and exothermic beyond it. Eq.\,(\ref{Eq:TNcontrolV}) exhibits its ease and effectiveness in guiding SOEC operational design. 

With Eq.\,(\ref{Eq:strategyB}), we can control $V_{\rm in}^{\rm fuel}$ to neutralize the heat source within the SOEC under variable solar conditions in numerical simulation. For example, when $G=200$\,W\,m$^{-2}$, the power applied on the single-channel SOEC model is 0.87\,W. Then, adjusting $V_{\rm in}^{\rm fuel}$ to $0.3$\,m/s as calculated using Eq.\,(\ref{Eq:strategyB}) could maintain the total heat source at around zero. As shown in Fig.\,\ref{fig:CFDResults}g, the adjustment on the $V_{\rm in}^{\rm fuel}$ leads to a highly uniform temperature distribution within the SOEC, indicating the effectiveness of the proposed strategy.

Alternatively, by adjusting the steam flow rate, SOEC can be maintained at slightly endothermicity for higher electrical efficiency. In this study, Strategy C is designed to maintain a slight temperature gradient of -3\,K\,cm$^{-1}$ within the SOEC, as shown in Fig.\,\ref{fig:deltaTV}. Such temperature gradient is acceptable as it falls within the suggested magnitude of 5\,K\,cm$^{-1}$~\cite{CHEN2023117596,WANG2019255}. To develop Strategy C, the relationship between the steam flow rate $V_{\rm in}^{\rm fuel}$ and the input power $P_{\rm SOEC,SCh}$ is regressed at $\Delta T_{\rm in,out}/L_{\rm SCh}=-3$\,K\,cm$^{-1}$, as presented in Eq.\,(\ref{Eq:strategyC}). 

\subsection{Controlling heat source in SOEC under fluctuating solar conditions}
\subsubsection*{Effectiveness of temperature control}
Figure~\ref{fig:GHI_time} shows the comparison of four different operating strategies for the SOEC under fluctuating solar irradiance. The one-hour GHI data with a temporal resolution of 2~s collected in Hong Kong from 11:59:10 to 12:59:10 on August 30th, 2023, serves as the input for the coupled PV-SOEC model. Transient simulations were carried out with a constant timestep of 2 seconds. The four operating strategies applied on the SOEC are detailed in Table\,\ref{Tab:BC}. The base case refers to a scenario with constant operating parameters. In Strategy A, the air flow rate is increased to five times that of the base case. Strategies B and C manipulate the steam velocity according to GHI using Eqs.\,(\ref{Eq:strategyB}-\ref{Eq:strategyC}), aiming to maintain SOEC at states of thermal neutrality and slight endothermicity (the temperature gradient of -3\,K\,cm$^{-1}$), respectively. All four cases are initialized with the same condition, which is the steady-state solution with the base case settings when $G= 1000$\,W\,m$^{-2}$.

\begin{figure}
    \centering
    \includegraphics[width=\textwidth]{figures/GHI_time_cloud.png}
\end{figure}
\begin{figure}
\caption{Comparison of four different operating strategies on the PV-SOEC system under fluctuating solar irradiance. (a) shows the one-hour GHI data and sky image. The variables for comparison includes: (b) outlet temperature and temperature gradient, (c) temperature variation rate, (d) the total heat source within the single-channel SOEC model, (e) electrical efficiency, and (f) current density and hydrogen production rate. }
    \label{fig:GHI_time}
\end{figure}

Figure~\ref{fig:GHI_time}a presents the measured GHI data and the corresponding sky images. During the study period, dynamic cloud fields result in significant fluctuations in GHI, ranging from about 1000 to 200\,W\,m$^{-2}$ within the hour. As shown in Fig.~\ref{fig:GHI_time}b, the oscillating solar irradiance induces a temperature gradient within the SOEC in the base case. This gradient exceeds the recommended limit of 5\,K\,cm$^{-1}$~\cite{CHEN2023117596,WANG2019255} during high solar variability periods. For Strategy A, the increase of air flow rate can reduce the temperature gradient to below the suggested limit. For Strategy B, by manipulating the steam velocity according to solar power, the temperature is regulated towards thermal neutrality and the temperature gradient is nearly eliminated. Strategy C also successfully maintains the SOEC at a slightly endothermic state with a safe temperature gradient of around -3\,K\,cm$^{-1}$.

Figure~\ref{fig:GHI_time}c compares the variation rates of the average temperature of SOEC in the four cases. The temperature variation rates in the base case and Strategy A are heavily influenced by GHI fluctuations and frequently exceed the recommended limit of 1\,K\,min$^{-1}$~\cite{WANG2019255}. These rates even surpass 5\,K\,min$^{-1}$ at certain times, such as at 12:07, 12:15, and 12:28. Such high temperature variation rates may induce thermal fatigue and increase the likelihood of SOEC failure. For Strategy B, the temperature variation rate exceeds 1\,K\,min$^{-1}$ within the initial 7 minutes due to an overshoot of heat source as shown in Fig.~\ref{fig:GHI_time}d. Afterwards, the SOEC operates safely and stably within the limit. Overall, in Strategies B and C, the control of steam velocity successfully reduces the temperature variation rate to below the suggested limit of 1\,K\,min$^{-1}$. 

The effectiveness of Strategies B and C is demonstrated by Figure~\ref{fig:GHI_time}d, which compares the total heat sources within the SOEC. For the base case, the heat source within SOEC is highly sensitive to the fluctuating solar power. large-amplitude variations in the heat source lead to rapid temperature changes. Strategy A is unable to effectively mitigate these fluctuations by merely increasing the air flow rate, thus failing to maintain the rate of temperature change within the recommended limits. In contrast, Strategies B and C achieve success in managing the heat source by adjusting the steam flow rate of SOEC. On one hand, they suppress the large-amplitude variations of the heat source, resulting in a reduced rate of temperature change. On the other hand, they keep the magnitude of the heat source at a low level, which contributes to a small temperature gradient within the SOEC.

In addition, the slow temperature regulation process of SOEC should be highlighted in Fig.~\ref{fig:GHI_time}b. Due to the regulation of Strategy B, the outlet temperature $T_{\rm out}$ of SOEC increases from the initial state of 1040\,K to the thermoneutral point of 1073\,K. The temperature increase lasts for approximately 20 minutes. Such a long time required for the temperature regulation of SOEC is also observed by \citet{AGUIAR2005136}. We propose that the considerable time for temperature regulation is associated with the heat-transfer characteristic time of the SOEC~\cite{liang2023characteristic}, which represents the time necessary for the SOEC to achieve a new thermal equilibrium after a variation of heat source. In our previous work~\cite{liang2023characteristic}, we have explained the physics of the long time for temperature regulation through Eq.\,(\ref{eq:tau_th}), which is comparable to the time required for heat fluxes to `fill' the heat capacity of SOEC. Eq.\,(\ref{eq:tau_th}) can be used to estimate the characteristic time of heat transfer,
\begin{equation}
\begin{split}
\tau_{\text h}&=\frac{\text { Total enthalpy }}{\text { Total heat transfer rate }} \frac{[\rm{J}]}{[\rm{W}]} = \frac{\mathcal{H}_{\rm{0,int}}^{\text {solid }}+\mathcal{H}_{\rm{0,E}}^{\rm{solid}}+\mathcal{H}_{\rm{0,AFL}}^{\rm{eff}}+\mathcal{H}_{\rm{0,ADL}}^{\rm{eff}}+\mathcal{H}_{\rm{0,CDL}}^{\rm{eff}}+\mathcal{H}_{\rm{0,CFL}}^{\rm{eff}}+\mathcal{H}_{\rm{0,ch}}^{\rm{fluid}}}{\dot{\mathcal{H}}_{\rm{in}}^{\rm{fuel}}+\dot{\mathcal{H}}_{\rm{in}}^{\rm{air}}}\\ 
& \approx \frac{(mc_p)_{\rm{0,int}}^{\text {solid }}+(mc_p)_{\rm{0,E}}^{\rm{solid}}+(mc_p)_{\rm{0,AFL}}^{\rm{eff}}+(mc_p)_{\rm{0,ADL}}^{\rm{eff}}+(mc_p)_{\rm{0,CDL}}^{\rm{eff}}+(mc_p)_{\rm{0,CFL}}^{\rm{eff}}+(mc_p)_{\rm{0,ch}}^{\rm{fluid}}}{(\dot{m}c_p)_{\rm{in}}^{\rm{fuel}}+(\dot{m}c_p)_{\rm{in}}^{\rm{air}}}
\end{split}\label{eq:tau_th}
\end{equation}
where, $\dot{\mathcal{H}}$ is the enthalpy flow rate, $m$ is mass and $\dot{m}$ is mass flow rate. The subscripts `E', `int', and `ch' correspond to the electrolyte, interconnect, and fluid channel, respectively. In Strategy B, the estimated characteristic time of heat transfer is 1100\,s, which closely aligns with the simulated temperature regulation time for Strategy B. This indicates a correlation between the slow temperature regulation process and the heat-transfer characteristic time of SOEC.

\subsubsection*{Comparison of efficiencies}
Although Strategy B exhibits superior temperature uniformity and thermal stability, it shows an electrical efficiency of 91.2\%, lower than the 100.7\% of base case and the 103.0\% of Case 1, as shown in Fig.\ref{fig:GHI_time}e. Interestingly, at low GHI, the efficiency of Strategy B decreases, while the efficiencies of the base case and Case 1 can increases to even higher than 100\%. This phenomenon is attributed to the energy utilization of the SOEC. The SOEC consumes both electrical and thermal energy for electrolysis. When GHI is low, the base case and Strategy A consume more thermal energy than Strategy C to compensate for the decrease in electrical energy, leading to improved electrical efficiency and stronger endothermic effects. The enhanced endothermicity unavoidably results in a greater temperature gradient. This situation reveals an inherent contradiction between thermal control and the maximization of efficiency. Strategy C presents a compromise to this dilemma by operating the SOEC at the slightly endothermic state, achieving an electrical efficiency of 94.9\%. In comparison to thermal neutrality, the slight endothermicity permits an acceptable temperature gradient within SOEC ($<5$\,K\,cm$^{-1}$) while maintains the high electrolysis efficiency.

Figure~\ref{fig:GHI_time}f provides an additional comparison of the currents and hydrogen production rate in the four cases. Based on Eq.\,(\ref{Eq:H2Rate}), the average hydrogen production rates for the base case and Strategies A-C  are  140.9\,L\,h$^{-1}$, 144.1\,L\,h$^{-1}$, 130.4\,L\,h$^{-1}$, and 135.5\,L\,h$^{-1}$. This outcome further illustrates that operating SOEC at slight endothermicity can effectively balance thermal management and system efficiency.

\section{Conclusion}
\label{Sec:conclusion}
To enhance the thermal safety of SOEC when integrated with PV panels under fluctuating solar irradiance, theoretical analysis and numerical simulation are conducted to investigate the control strategies for SOEC. Recognizing that the maximum current of SOEC is intrinsically limited by steam supply rate, we developed control strategies that adaptively adjust the steam velocity according to real-time solar power. In this way, the magnitude of heat source within SOEC can be maintained at a low level, thereby ensuring that the temperature gradient and variation within SOEC remain below 5\,K\,cm$^{-1}$ and 1\,K\,min$^{-1}$, respectively, even under highly fluctuating solar power. 
This would enhance the durability and safety of SOEC when integrated with intermittent power sources. 

In addition, we found that while the thermoneutral operation of SOEC shows superior temperature uniformity and stability, it lags behind the endothermic operations in terms of electrical efficiency, as it lacks the utilization of thermal energy to compensate for electrical energy. This observation indicates a contradiction between temperature uniformity and the maximization of efficiency in SOEC. To strike a balance for such contradiction, our control strategy can alternatively maintain the SOEC at a slight endothermic state for a high efficiency of 94.9\% and an acceptable temperature gradient below 5\,K\,cm$^{-1}$ under fluctuating solar power. 

\section*{Acknowledgement} 
The authors gratefully acknowledge the partial support from The Hong Kong Polytechnic University (P0035016), National Natural Science Foundation of China (No. 52306236), Science, Technology and Innovation Commission of Shenzhen Municipality (GXWD20220811165757005), and Department of Education of Guangdong Province (2021KQNCX271). 

\section*{Declaration of generative AI and AI-assisted technologies in the writing process}
During the preparation of this work, the authors used GPT-4 in order to improve language and readability. After using this tool, the authors reviewed and edited the content as needed and take full responsibility for the content of the publication.

\mbox{}
\nomenclature[A]{$\delta$}{Thickness, [m]}
\nomenclature[A]{$\Delta T$}{Temperature difference, [K]}
\nomenclature[A]{$T$}{Temperature, [K]}
\nomenclature[A]{$U$}{Voltage, [V]}
\nomenclature[A]{$\phi$}{Electrical potential, [V]}
\nomenclature[A]{$i$}{Current density, [A\,m$^{-2}$]}
\nomenclature[A]{$I$}{Current, [A]}
\nomenclature[A]{$R$}{Electrical resistance, [$\Omega$]}
\nomenclature[A]{$q$}{Elementary charge, [C]}
\nomenclature[A]{$k_{\rm B}$}{Boltzmann constant, [J\,K$^{-1}$]}
\nomenclature[A]{$G$}{Global horizontal irradiance, [W\,m$^{-2}$]}
\nomenclature[A]{$\mu$}{Short-circuit current temperature coefficient, [A\,K$^{-1}$]}
\nomenclature[A]{$E_{\rm{g}}$}{Silicon-based PV panel band-gap, [eV]}
\nomenclature[A]{$P$}{Power, [W]}
\nomenclature[A]{$p$}{Pressure, [Pa]}
\nomenclature[A]{$X_i$}{Mole fraction of species $i$}
\nomenclature[A]{$A_{\rm act}$}{Active area, [m$^{-2}$]}
\nomenclature[A]{$A_{\rm ch}$}{Cross area of fluid channel, [m$^{-2}$]}
\nomenclature[A]{$\dot{\mathcal{V}}$}{Volumetric flow rate, [m$^3$\,s$^{-1}$]}
\nomenclature[A]{$V$}{Velocity, [m\,s$^{-1}$]}
\nomenclature[A]{$n$}{Number of participating electrons for hydrogen production}
\nomenclature[A]{$n_{\rm s}$}{Number of devices connected in series}
\nomenclature[A]{$n_{\rm p}$}{Number of devices connected in parallel}
\nomenclature[A]{$V$}{Velocity, [m\,s$^{-1}$]}
\nomenclature[A]{$\Delta G$}{Gibbs free energy change of water formation, [${\rm J\,mol^{-1}}$]}
\nomenclature[A]{$F$}{Faraday constant, [${\rm C\,mol^{-1}}$]}
\nomenclature[A]{$\dot{Q}_{\rm H2}$}{Hydrogen production rate, [L\,h$^{-1}$]}
\nomenclature[A]{$M_i$}{Molar mass of species $i$, [kg\,mol$^{-1}$]}
\nomenclature[A]{$\Delta t$}{Time step size, [s]}
\nomenclature[A]{$\rho$}{Density, [kg\,m$^{-3}$]}
\nomenclature[A]{$L$}{Length, [m]}
\nomenclature[A]{$W$}{Width, [m]}
\nomenclature[A]{$\dot{\mathcal{H}}$}{Enthalpy flux, [W]}
\nomenclature[A]{$\mathcal{H}$}{Enthalpy, [J]}
\nomenclature[A]{$\mathcal{R}$}{Ideal gas constant, [J\,mol$^{-1}$\,K$^{-1}$]}
\nomenclature[A]{$S_{\rm H}$}{Heat source, [W]}
\nomenclature[A]{$S_{\rm Ohm}$}{Ohmic heat source, [W]}
\nomenclature[A]{$S_{\rm op}$}{Overpotential heat source, [W]}
\nomenclature[A]{$S_{\rm rev}$}{Reversible heat source, [W]}
\nomenclature[A]{$\eta$}{Overpotential, [V]}
\nomenclature[A]{$c_p$}{Specific heat, [J\,kg$^{-1}$\,K$^{-1}$]}
\nomenclature[A]{$\Delta\mathcal{S}_{\rm f}$}{Entropy of formation of water, [J\,mol$^{-1}$\,K$^{-1}$]}
\nomenclature[A]{$\Delta\mathcal{H}_{\rm f}$}{Enthalpy of formation of water, [J\,mol$^{-1}$]}
\nomenclature[A]{$m$}{Mass, [kg]}
\nomenclature[A]{$\dot{m}$}{Mass flow rate, [kg\,s$^{-1}$]}
\nomenclature[A]{$t$}{Time, [s]}
\nomenclature[A]{$\tau_{\rm h}$}{Heat-transfer time constant, [s]}

\nomenclature[B]{s}{Series}
\nomenclature[B]{p}{Parallel}
\nomenclature[B]{SCh}{Single-channel model}
\nomenclature[B]{ele}{Electronic}
\nomenclature[B]{act}{Active}
\nomenclature[B]{ave}{Average}
\nomenclature[B]{TN}{Thermal neutrality}
\nomenclature[B]{lim}{Limit}
\nomenclature[B]{ch}{Channel}
\nomenclature[B]{int}{Interconnect}
\nomenclature[B]{E}{Electrolyte}

\nomenclature[C]{FL}{Functional layer}
\nomenclature[C]{theo}{Theoretical}
\nomenclature[C]{eff}{Effective}

\nomenclature[D]{SOEC}{Solid oxide electrolysis cell}
\nomenclature[D]{SOFC}{Solid oxide fuel cell}
\nomenclature[D]{MPP}{Maximum power point}
\nomenclature[D]{BC}{Boundary condition}
\nomenclature[D]{GHI}{Global horizontal irradiance}
\nomenclature[D]{PV}{Photovoltaic}
\nomenclature[D]{AFL}{Anode functional layer}
\nomenclature[D]{CFL}{Cathode functional layer}
\nomenclature[D]{ADL}{Anode diffusion layer}
\nomenclature[D]{CDL}{Cathode diffusion layer}
\nomenclature[D]{UDF}{User Defined Function}
\printnomenclature

\bibliographystyle{elsarticle-num-names} 
\biboptions{sort&compress} 
\bibliography{Refs} 

\begin{thebibliography}{42}
\expandafter\ifx\csname natexlab\endcsname\relax\def\natexlab#1{#1}\fi
\providecommand{\url}[1]{\texttt{#1}}
\providecommand{\href}[2]{#2}
\providecommand{\path}[1]{#1}
\providecommand{\DOIprefix}{doi:}
\providecommand{\ArXivprefix}{arXiv:}
\providecommand{\URLprefix}{URL: }
\providecommand{\Pubmedprefix}{pmid:}
\providecommand{\doi}[1]{\href{http://dx.doi.org/#1}{\path{#1}}}
\providecommand{\Pubmed}[1]{\href{pmid:#1}{\path{#1}}}
\providecommand{\bibinfo}[2]{#2}
\ifx\xfnm\relax \def\xfnm[#1]{\unskip,\space#1}\fi
\bibitem[{Wang et~al.(2023)Wang, Wu, Zhao, Guo, Han, Zhao, Zu, Du, Ni, and Jiao}]{WANG2023516}
\bibinfo{author}{Y.~Wang}, \bibinfo{author}{C.~Wu}, \bibinfo{author}{S.~Zhao}, \bibinfo{author}{Z.~Guo}, \bibinfo{author}{M.~Han}, \bibinfo{author}{T.~Zhao}, \bibinfo{author}{B.~Zu}, \bibinfo{author}{Q.~Du}, \bibinfo{author}{M.~Ni}, \bibinfo{author}{K.~Jiao},
\newblock \bibinfo{title}{Boosting the performance and durability of heterogeneous electrodes for solid oxide electrochemical cells utilizing a data-driven powder-to-power framework},
\newblock \bibinfo{journal}{Science Bulletin} \bibinfo{volume}{68} (\bibinfo{year}{2023}) \bibinfo{pages}{516--527}. \DOIprefix\doi{https://doi.org/10.1016/j.scib.2023.02.019}.
\bibitem[{IEA(2023)}]{iea2023world}
\bibinfo{author}{IEA},
\newblock \bibinfo{title}{World energy outlook 2023},
\newblock \bibinfo{organization}{IEA Paris, France}, \bibinfo{year}{2023}. \URLprefix \url{https://www.iea.org/reports/world-energy-outlook-2023}.
\bibitem[{Hu et~al.(2022)Hu, Fang, Ai, Huang, Zhong, Yang, and Wang}]{HU2022118788}
\bibinfo{author}{K.~Hu}, \bibinfo{author}{J.~Fang}, \bibinfo{author}{X.~Ai}, \bibinfo{author}{D.~Huang}, \bibinfo{author}{Z.~Zhong}, \bibinfo{author}{X.~Yang}, \bibinfo{author}{L.~Wang},
\newblock \bibinfo{title}{Comparative study of alkaline water electrolysis, proton exchange membrane water electrolysis and solid oxide electrolysis through multiphysics modeling},
\newblock \bibinfo{journal}{Applied Energy} \bibinfo{volume}{312} (\bibinfo{year}{2022}) \bibinfo{pages}{118788}. \DOIprefix\doi{https://doi.org/10.1016/j.apenergy.2022.118788}.
\bibitem[{Min et~al.(2022)Min, Choi, and Hong}]{MIN2022120145}
\bibinfo{author}{G.~Min}, \bibinfo{author}{S.~Choi}, \bibinfo{author}{J.~Hong},
\newblock \bibinfo{title}{A review of solid oxide steam-electrolysis cell systems: Thermodynamics and thermal integration},
\newblock \bibinfo{journal}{Applied Energy} \bibinfo{volume}{328} (\bibinfo{year}{2022}) \bibinfo{pages}{120145}. \DOIprefix\doi{https://doi.org/10.1016/j.apenergy.2022.120145}.
\bibitem[{Xu et~al.(2022)Xu, Guo, Xia, He, Li, {Temitope Bello}, Zheng, and Ni}]{XU2022115175}
\bibinfo{author}{Q.~Xu}, \bibinfo{author}{Z.~Guo}, \bibinfo{author}{L.~Xia}, \bibinfo{author}{Q.~He}, \bibinfo{author}{Z.~Li}, \bibinfo{author}{I.~{Temitope Bello}}, \bibinfo{author}{K.~Zheng}, \bibinfo{author}{M.~Ni},
\newblock \bibinfo{title}{A comprehensive review of solid oxide fuel cells operating on various promising alternative fuels},
\newblock \bibinfo{journal}{Energy Conversion and Management} \bibinfo{volume}{253} (\bibinfo{year}{2022}) \bibinfo{pages}{115175}. \DOIprefix\doi{https://doi.org/10.1016/j.enconman.2021.115175}.
\bibitem[{Beale et~al.(2021)Beale, Andersson, Boigues-Muñoz, Frandsen, Lin, McPhail, Ni, Sundén, Weber, and Weber}]{BEALE2021100902}
\bibinfo{author}{S.~B. Beale}, \bibinfo{author}{M.~Andersson}, \bibinfo{author}{C.~Boigues-Muñoz}, \bibinfo{author}{H.~L. Frandsen}, \bibinfo{author}{Z.~Lin}, \bibinfo{author}{S.~J. McPhail}, \bibinfo{author}{M.~Ni}, \bibinfo{author}{B.~Sundén}, \bibinfo{author}{A.~Weber}, \bibinfo{author}{A.~Z. Weber},
\newblock \bibinfo{title}{Continuum scale modelling and complementary experimentation of solid oxide cells},
\newblock \bibinfo{journal}{Progress in Energy and Combustion Science} \bibinfo{volume}{85} (\bibinfo{year}{2021}) \bibinfo{pages}{100902}. \DOIprefix\doi{https://doi.org/10.1016/j.pecs.2020.100902}.
\bibitem[{Petipas et~al.(2013)Petipas, Fu, Brisse, and Bouallou}]{PETIPAS20132957}
\bibinfo{author}{F.~Petipas}, \bibinfo{author}{Q.~Fu}, \bibinfo{author}{A.~Brisse}, \bibinfo{author}{C.~Bouallou},
\newblock \bibinfo{title}{Transient operation of a solid oxide electrolysis cell},
\newblock \bibinfo{journal}{International Journal of Hydrogen Energy} \bibinfo{volume}{38} (\bibinfo{year}{2013}) \bibinfo{pages}{2957--2964}. \DOIprefix\doi{https://doi.org/10.1016/j.ijhydene.2012.12.086}.
\bibitem[{Zeng et~al.(2020)Zeng, Qian, Zhang, Hao, Dan, and Zhuge}]{ZENG2020115899}
\bibinfo{author}{Z.~Zeng}, \bibinfo{author}{Y.~Qian}, \bibinfo{author}{Y.~Zhang}, \bibinfo{author}{C.~Hao}, \bibinfo{author}{D.~Dan}, \bibinfo{author}{W.~Zhuge},
\newblock \bibinfo{title}{A review of heat transfer and thermal management methods for temperature gradient reduction in solid oxide fuel cell (sofc) stacks},
\newblock \bibinfo{journal}{Applied Energy} \bibinfo{volume}{280} (\bibinfo{year}{2020}) \bibinfo{pages}{115899}. \DOIprefix\doi{https://doi.org/10.1016/j.apenergy.2020.115899}.
\bibitem[{Wang et~al.(2019)Wang, Banerjee, and Deutschmann}]{WANG2019255}
\bibinfo{author}{Y.~Wang}, \bibinfo{author}{A.~Banerjee}, \bibinfo{author}{O.~Deutschmann},
\newblock \bibinfo{title}{Dynamic behavior and control strategy study of co2/h2o co-electrolysis in solid oxide electrolysis cells},
\newblock \bibinfo{journal}{Journal of Power Sources} \bibinfo{volume}{412} (\bibinfo{year}{2019}) \bibinfo{pages}{255--264}. \DOIprefix\doi{https://doi.org/10.1016/j.jpowsour.2018.11.047}.
\bibitem[{Sun et~al.(2022)Sun, Lu, Liu, Shuai, Sun, Zheng, Han, Xiao, Xuan, Ni, and Xu}]{SUN2022115560}
\bibinfo{author}{Y.~Sun}, \bibinfo{author}{J.~Lu}, \bibinfo{author}{Q.~Liu}, \bibinfo{author}{W.~Shuai}, \bibinfo{author}{A.~Sun}, \bibinfo{author}{N.~Zheng}, \bibinfo{author}{Y.~Han}, \bibinfo{author}{G.~Xiao}, \bibinfo{author}{J.~Xuan}, \bibinfo{author}{M.~Ni}, \bibinfo{author}{H.~Xu},
\newblock \bibinfo{title}{Multi-objective optimizations of solid oxide co-electrolysis with intermittent renewable power supply via multi-physics simulation and deep learning strategy},
\newblock \bibinfo{journal}{Energy Conversion and Management} \bibinfo{volume}{258} (\bibinfo{year}{2022}) \bibinfo{pages}{115560}. \DOIprefix\doi{https://doi.org/10.1016/j.enconman.2022.115560}.
\bibitem[{Xiao et~al.(2023)Xiao, Sun, Liu, Ni, and Xu}]{XIAO2023120383}
\bibinfo{author}{G.~Xiao}, \bibinfo{author}{A.~Sun}, \bibinfo{author}{H.~Liu}, \bibinfo{author}{M.~Ni}, \bibinfo{author}{H.~Xu},
\newblock \bibinfo{title}{Thermal management of reversible solid oxide cells in the dynamic mode switching},
\newblock \bibinfo{journal}{Applied Energy} \bibinfo{volume}{331} (\bibinfo{year}{2023}) \bibinfo{pages}{120383}. \DOIprefix\doi{https://doi.org/10.1016/j.apenergy.2022.120383}.
\bibitem[{Liu et~al.(2022)Liu, Zhao, Li, Xia, Jiang, Kupecki, Pang, Deng, and Li}]{LIU2022115318}
\bibinfo{author}{G.~Liu}, \bibinfo{author}{W.~Zhao}, \bibinfo{author}{Z.~Li}, \bibinfo{author}{Z.~Xia}, \bibinfo{author}{C.~Jiang}, \bibinfo{author}{J.~Kupecki}, \bibinfo{author}{S.~Pang}, \bibinfo{author}{Z.~Deng}, \bibinfo{author}{X.~Li},
\newblock \bibinfo{title}{Modeling and control-oriented thermal safety analysis for mode switching process of reversible solid oxide cell system},
\newblock \bibinfo{journal}{Energy Conversion and Management} \bibinfo{volume}{255} (\bibinfo{year}{2022}) \bibinfo{pages}{115318}. \DOIprefix\doi{https://doi.org/10.1016/j.enconman.2022.115318}.
\bibitem[{Udagawa et~al.(2008)Udagawa, Aguiar, and Brandon}]{UDAGAWA200846}
\bibinfo{author}{J.~Udagawa}, \bibinfo{author}{P.~Aguiar}, \bibinfo{author}{N.~Brandon},
\newblock \bibinfo{title}{Hydrogen production through steam electrolysis: Model-based dynamic behaviour of a cathode-supported intermediate temperature solid oxide electrolysis cell},
\newblock \bibinfo{journal}{Journal of Power Sources} \bibinfo{volume}{180} (\bibinfo{year}{2008}) \bibinfo{pages}{46--55}. \DOIprefix\doi{https://doi.org/10.1016/j.jpowsour.2008.02.026}.
\bibitem[{Zhang et~al.(2022)Zhang, Maloney, {Farida Harun}, Zhou, Pezzini, Medam, Hovsapian, Bayham, and Tucker}]{ZHANG2022115544}
\bibinfo{author}{B.~Zhang}, \bibinfo{author}{D.~Maloney}, \bibinfo{author}{N.~{Farida Harun}}, \bibinfo{author}{N.~Zhou}, \bibinfo{author}{P.~Pezzini}, \bibinfo{author}{A.~Medam}, \bibinfo{author}{R.~Hovsapian}, \bibinfo{author}{S.~Bayham}, \bibinfo{author}{D.~Tucker},
\newblock \bibinfo{title}{Rapid load transition for integrated solid oxide fuel cell – gas turbine (sofc-gt) energy systems: A demonstration of the potential for grid response},
\newblock \bibinfo{journal}{Energy Conversion and Management} \bibinfo{volume}{258} (\bibinfo{year}{2022}) \bibinfo{pages}{115544}. \DOIprefix\doi{https://doi.org/10.1016/j.enconman.2022.115544}.
\bibitem[{Zhu et~al.(2023)Zhu, Wu, Yang, Wang, Li, Yang, and Zhang}]{ZHU2023121655}
\bibinfo{author}{P.~Zhu}, \bibinfo{author}{Z.~Wu}, \bibinfo{author}{Y.~Yang}, \bibinfo{author}{H.~Wang}, \bibinfo{author}{R.~Li}, \bibinfo{author}{F.~Yang}, \bibinfo{author}{Z.~Zhang},
\newblock \bibinfo{title}{The dynamic response of solid oxide fuel cell fueled by syngas during the operating condition variations},
\newblock \bibinfo{journal}{Applied Energy} \bibinfo{volume}{349} (\bibinfo{year}{2023}) \bibinfo{pages}{121655}. \DOIprefix\doi{https://doi.org/10.1016/j.apenergy.2023.121655}.
\bibitem[{Botta et~al.(2019)Botta, Romeo, Fernandes, Trabucchi, and Aravind}]{BOTTA2019636}
\bibinfo{author}{G.~Botta}, \bibinfo{author}{M.~Romeo}, \bibinfo{author}{A.~Fernandes}, \bibinfo{author}{S.~Trabucchi}, \bibinfo{author}{P.~Aravind},
\newblock \bibinfo{title}{Dynamic modeling of reversible solid oxide cell stack and control strategy development},
\newblock \bibinfo{journal}{Energy Conversion and Management} \bibinfo{volume}{185} (\bibinfo{year}{2019}) \bibinfo{pages}{636--653}. \DOIprefix\doi{https://doi.org/10.1016/j.enconman.2019.01.082}.
\bibitem[{Aguiar et~al.(2005)Aguiar, Adjiman, and Brandon}]{AGUIAR2005136}
\bibinfo{author}{P.~Aguiar}, \bibinfo{author}{C.~Adjiman}, \bibinfo{author}{N.~Brandon},
\newblock \bibinfo{title}{Anode-supported intermediate-temperature direct internal reforming solid oxide fuel cell: Ii. model-based dynamic performance and control},
\newblock \bibinfo{journal}{Journal of Power Sources} \bibinfo{volume}{147} (\bibinfo{year}{2005}) \bibinfo{pages}{136--147}. \DOIprefix\doi{https://doi.org/10.1016/j.jpowsour.2005.01.017}.
\bibitem[{Lu et~al.(2024)Lu, Zhang, Zhang, Zhang, Zhu, and Huang}]{LU2024117852}
\bibinfo{author}{B.~Lu}, \bibinfo{author}{Z.~Zhang}, \bibinfo{author}{Z.~Zhang}, \bibinfo{author}{C.~Zhang}, \bibinfo{author}{L.~Zhu}, \bibinfo{author}{Z.~Huang},
\newblock \bibinfo{title}{Control strategy of solid oxide electrolysis cell operating temperature under real fluctuating renewable power},
\newblock \bibinfo{journal}{Energy Conversion and Management} \bibinfo{volume}{299} (\bibinfo{year}{2024}) \bibinfo{pages}{117852}. \DOIprefix\doi{https://doi.org/10.1016/j.enconman.2023.117852}.
\bibitem[{Liu et~al.(2023)Liu, Wang, Liu, Kupecki, Zhao, Jin, Wang, and Li}]{LIU2023139000}
\bibinfo{author}{G.~Liu}, \bibinfo{author}{Z.~Wang}, \bibinfo{author}{X.~Liu}, \bibinfo{author}{J.~Kupecki}, \bibinfo{author}{D.~Zhao}, \bibinfo{author}{B.~Jin}, \bibinfo{author}{Z.~Wang}, \bibinfo{author}{X.~Li},
\newblock \bibinfo{title}{Transient analysis and safety-oriented process optimization during electrolysis–fuel cell transition of a novel reversible solid oxide cell system},
\newblock \bibinfo{journal}{Journal of Cleaner Production} \bibinfo{volume}{425} (\bibinfo{year}{2023}) \bibinfo{pages}{139000}. \DOIprefix\doi{https://doi.org/10.1016/j.jclepro.2023.139000}.
\bibitem[{Chen et~al.(2023)Chen, Wu, Luo, Wang, and Xu}]{CHEN2023117596}
\bibinfo{author}{H.~Chen}, \bibinfo{author}{T.~Wu}, \bibinfo{author}{S.~Luo}, \bibinfo{author}{Y.~Wang}, \bibinfo{author}{X.~Xu},
\newblock \bibinfo{title}{Operating strategy investigation of a solid oxide electrolysis cell under large scale transient electrical inputs},
\newblock \bibinfo{journal}{Energy Conversion and Management} \bibinfo{volume}{294} (\bibinfo{year}{2023}) \bibinfo{pages}{117596}. \DOIprefix\doi{https://doi.org/10.1016/j.enconman.2023.117596}.
\bibitem[{Tseronis et~al.(2012)Tseronis, Bonis, Kookos, and Theodoropoulos}]{TSERONIS2012530}
\bibinfo{author}{K.~Tseronis}, \bibinfo{author}{I.~Bonis}, \bibinfo{author}{I.~Kookos}, \bibinfo{author}{C.~Theodoropoulos},
\newblock \bibinfo{title}{Parametric and transient analysis of non-isothermal, planar solid oxide fuel cells},
\newblock \bibinfo{journal}{International Journal of Hydrogen Energy} \bibinfo{volume}{37} (\bibinfo{year}{2012}) \bibinfo{pages}{530--547}. \DOIprefix\doi{https://doi.org/10.1016/j.ijhydene.2011.09.062}, \bibinfo{note}{11th China Hydrogen Energy Conference}.
\bibitem[{Srikanth et~al.(2018)Srikanth, Heddrich, Gupta, and Friedrich}]{SRIKANTH2018473}
\bibinfo{author}{S.~Srikanth}, \bibinfo{author}{M.~Heddrich}, \bibinfo{author}{S.~Gupta}, \bibinfo{author}{K.~Friedrich},
\newblock \bibinfo{title}{Transient reversible solid oxide cell reactor operation – experimentally validated modeling and analysis},
\newblock \bibinfo{journal}{Applied Energy} \bibinfo{volume}{232} (\bibinfo{year}{2018}) \bibinfo{pages}{473--488}. \DOIprefix\doi{https://doi.org/10.1016/j.apenergy.2018.09.186}.
\bibitem[{Sun et~al.(2022)Sun, Shuai, Zheng, Han, Xiao, Ni, and Xu}]{SUN2022116310}
\bibinfo{author}{A.~Sun}, \bibinfo{author}{W.~Shuai}, \bibinfo{author}{N.~Zheng}, \bibinfo{author}{Y.~Han}, \bibinfo{author}{G.~Xiao}, \bibinfo{author}{M.~Ni}, \bibinfo{author}{H.~Xu},
\newblock \bibinfo{title}{Self-adaptive heat management of solid oxide electrolyzer cell under fluctuating power supply},
\newblock \bibinfo{journal}{Energy Conversion and Management} \bibinfo{volume}{271} (\bibinfo{year}{2022}) \bibinfo{pages}{116310}. \DOIprefix\doi{https://doi.org/10.1016/j.enconman.2022.116310}.
\bibitem[{Tang et~al.(2019)Tang, Amiri, and Tad{\'{e}}}]{Tang2019}
\bibinfo{author}{S.~Tang}, \bibinfo{author}{A.~Amiri}, \bibinfo{author}{M.~O. Tad{\'{e}}},
\newblock \bibinfo{title}{{System Level Exergy Assessment of Strategies Deployed for Solid Oxide Fuel Cell Stack Temperature Regulation and Thermal Gradient Reduction}},
\newblock \bibinfo{journal}{Industrial and Engineering Chemistry Research} \bibinfo{volume}{58} (\bibinfo{year}{2019}) \bibinfo{pages}{2258--2267}. \DOIprefix\doi{10.1021/acs.iecr.8b04142}.
\bibitem[{Sun et~al.(2022)Sun, Zheng, Ji, Sun, Shuai, Zheng, Han, Xiao, Ni, and Xu}]{SUN2022102344}
\bibinfo{author}{Y.~Sun}, \bibinfo{author}{W.~Zheng}, \bibinfo{author}{S.~Ji}, \bibinfo{author}{A.~Sun}, \bibinfo{author}{W.~Shuai}, \bibinfo{author}{N.~Zheng}, \bibinfo{author}{Y.~Han}, \bibinfo{author}{G.~Xiao}, \bibinfo{author}{M.~Ni}, \bibinfo{author}{H.~Xu},
\newblock \bibinfo{title}{Dynamic behavior of high-temperature co2/h2o co-electrolysis coupled with real fluctuating renewable power},
\newblock \bibinfo{journal}{Sustainable Energy Technologies and Assessments} \bibinfo{volume}{52} (\bibinfo{year}{2022}) \bibinfo{pages}{102344}. \DOIprefix\doi{https://doi.org/10.1016/j.seta.2022.102344}.
\bibitem[{Yang et~al.(2021)Yang, Tong, Hauch, Sun, Yang, Peng, and Chen}]{YANG2021129260}
\bibinfo{author}{Y.~Yang}, \bibinfo{author}{X.~Tong}, \bibinfo{author}{A.~Hauch}, \bibinfo{author}{X.~Sun}, \bibinfo{author}{Z.~Yang}, \bibinfo{author}{S.~Peng}, \bibinfo{author}{M.~Chen},
\newblock \bibinfo{title}{Study of solid oxide electrolysis cells operated in potentiostatic mode: Effect of operating temperature on durability},
\newblock \bibinfo{journal}{Chemical Engineering Journal} \bibinfo{volume}{417} (\bibinfo{year}{2021}) \bibinfo{pages}{129260}. \DOIprefix\doi{https://doi.org/10.1016/j.cej.2021.129260}.
\bibitem[{Chen et~al.(2017)Chen, Sun, Chatzichristodoulou, Koch, Hendriksen, and Mogensen}]{Chen2017}
\bibinfo{author}{M.~Chen}, \bibinfo{author}{X.~Sun}, \bibinfo{author}{C.~Chatzichristodoulou}, \bibinfo{author}{S.~Koch}, \bibinfo{author}{P.~V. Hendriksen}, \bibinfo{author}{M.~B. Mogensen},
\newblock \bibinfo{title}{{Thermoneutral Operation of Solid Oxide Electrolysis Cells in Potentiostatic Mode}},
\newblock \bibinfo{journal}{ECS Meeting Abstracts} \bibinfo{volume}{MA2017-03} (\bibinfo{year}{2017}) \bibinfo{pages}{287--287}. \DOIprefix\doi{10.1149/ma2017-03/1/287}.
\bibitem[{Liao et~al.(2021)Liao, He, Xu, Dai, Cheng, and Ni}]{LIAO2021119798}
\bibinfo{author}{T.~Liao}, \bibinfo{author}{Q.~He}, \bibinfo{author}{Q.~Xu}, \bibinfo{author}{Y.~Dai}, \bibinfo{author}{C.~Cheng}, \bibinfo{author}{M.~Ni},
\newblock \bibinfo{title}{Coupling properties and parametric optimization of a photovoltaic panel driven thermoelectric refrigerators system},
\newblock \bibinfo{journal}{Energy} \bibinfo{volume}{220} (\bibinfo{year}{2021}) \bibinfo{pages}{119798}. \DOIprefix\doi{https://doi.org/10.1016/j.energy.2021.119798}.
\bibitem[{macrovector(2021)}]{hydrogenset}
\bibinfo{author}{macrovector}, \bibinfo{title}{Hydrogen energy set}, \bibinfo{year}{2021}. \URLprefix \url{https://www.freepik.com/free-vector/hydrogen-energy-set\_26762033.htm}.
\bibitem[{Liang et~al.(2023{\natexlab{a}})Liang, Wang, Wang, Ni, and Li}]{LIANG2023116759}
\bibinfo{author}{Z.~Liang}, \bibinfo{author}{J.~Wang}, \bibinfo{author}{Y.~Wang}, \bibinfo{author}{M.~Ni}, \bibinfo{author}{M.~Li},
\newblock \bibinfo{title}{Transient characteristics of a solid oxide electrolysis cell under different voltage ramps: Transport phenomena behind overshoots},
\newblock \bibinfo{journal}{Energy Conversion and Management} \bibinfo{volume}{279} (\bibinfo{year}{2023}{\natexlab{a}}) \bibinfo{pages}{116759}. \DOIprefix\doi{https://doi.org/10.1016/j.enconman.2023.116759}.
\bibitem[{Liang et~al.(2023{\natexlab{b}})Liang, Wang, An, Wang, Ni, and Li}]{liang2023characteristic}
\bibinfo{author}{Z.~Liang}, \bibinfo{author}{J.~Wang}, \bibinfo{author}{L.~An}, \bibinfo{author}{Y.~Wang}, \bibinfo{author}{M.~Ni}, \bibinfo{author}{M.~Li}, \bibinfo{title}{Characteristic time of transient response of solid oxide cells (socs) to changes in voltage/current: from theory to applications}, \bibinfo{year}{2023}{\natexlab{b}}. \href{http://arxiv.org/abs/2305.07926}{{\tt arXiv:2305.07926}}.
\bibitem[{PVp(????)}]{PVpanel}
???? \URLprefix \url{https://unboundsolar.com/1524436/suniva/solar-panels/suniva-opt285-60-4-100-silver-mono-solar-panel}.
\bibitem[{Bae et~al.(2019)Bae, Lee, and Hong}]{BAE2019112152}
\bibinfo{author}{Y.~Bae}, \bibinfo{author}{S.~Lee}, \bibinfo{author}{J.~Hong},
\newblock \bibinfo{title}{The effect of anode microstructure and fuel utilization on current relaxation and concentration polarization of solid oxide fuel cell under electrical load change},
\newblock \bibinfo{journal}{Energy Convers Manage} \bibinfo{volume}{201} (\bibinfo{year}{2019}) \bibinfo{pages}{112152}. \DOIprefix\doi{https://doi.org/10.1016/j.enconman.2019.112152}.
\bibitem[{Su et~al.(2022)Su, Zhong, and Jiao}]{Su2022}
\bibinfo{author}{Y.~Su}, \bibinfo{author}{Z.~Zhong}, \bibinfo{author}{Z.~Jiao},
\newblock \bibinfo{title}{{A novel multi-physics coupled heterogeneous single-cell numerical model for solid oxide fuel cell based on 3D microstructure reconstructions}},
\newblock \bibinfo{journal}{Energy and Environmental Science} \bibinfo{volume}{15} (\bibinfo{year}{2022}) \bibinfo{pages}{2410--2424}. \DOIprefix\doi{10.1039/d2ee00485b}.
\bibitem[{NIST(2021)}]{nist}
\bibinfo{author}{NIST}, \bibinfo{title}{Thermophysical properties of fluid systems}, \bibinfo{year}{2021}. \URLprefix \url{https://webbook.nist.gov/chemistry/fluid/}.
\bibitem[{Mas et~al.(2022)Mas, Berastain, Antoniou, Angeles, Valencia, and Celis}]{MAS2022116213}
\bibinfo{author}{R.~Mas}, \bibinfo{author}{A.~Berastain}, \bibinfo{author}{A.~Antoniou}, \bibinfo{author}{L.~Angeles}, \bibinfo{author}{S.~Valencia}, \bibinfo{author}{C.~Celis},
\newblock \bibinfo{title}{Genetic algorithms-based size optimization of directly and indirectly coupled photovoltaic-electrolyzer systems},
\newblock \bibinfo{journal}{Energy Conversion and Management} \bibinfo{volume}{270} (\bibinfo{year}{2022}) \bibinfo{pages}{116213}. \DOIprefix\doi{https://doi.org/10.1016/j.enconman.2022.116213}.
\bibitem[{Lin et~al.(2022)Lin, Suter, Diethelm, Van~herle, and Haussener}]{Lin2022}
\bibinfo{author}{M.~Lin}, \bibinfo{author}{C.~Suter}, \bibinfo{author}{S.~Diethelm}, \bibinfo{author}{J.~Van~herle}, \bibinfo{author}{S.~Haussener},
\newblock \bibinfo{title}{{Integrated solar-driven high-temperature electrolysis operating with concentrated irradiation}},
\newblock \bibinfo{journal}{Joule} \bibinfo{volume}{6} (\bibinfo{year}{2022}) \bibinfo{pages}{2102--2121}. \DOIprefix\doi{10.1016/j.joule.2022.07.013}.
\bibitem[{Tembhurne et~al.(2019)Tembhurne, Nandjou, and Haussener}]{Tembhurne2019}
\bibinfo{author}{S.~Tembhurne}, \bibinfo{author}{F.~Nandjou}, \bibinfo{author}{S.~Haussener},
\newblock \bibinfo{title}{{A thermally synergistic photo-electrochemical hydrogen generator operating under concentrated solar irradiation}},
\newblock \bibinfo{journal}{Nature Energy} \bibinfo{volume}{4} (\bibinfo{year}{2019}) \bibinfo{pages}{399--407}. \DOIprefix\doi{10.1038/s41560-019-0373-7}.
\bibitem[{Bard et~al.(2022)Bard, Faulkner, and White}]{bard2022}
\bibinfo{author}{A.~J. Bard}, \bibinfo{author}{L.~R. Faulkner}, \bibinfo{author}{H.~S. White}, \bibinfo{title}{Electrochemical methods: fundamentals and applications}, \bibinfo{publisher}{John Wiley \& Sons}, \bibinfo{year}{2022}.
\bibitem[{Paul and Andrews(2008)}]{PAUL2008490}
\bibinfo{author}{B.~Paul}, \bibinfo{author}{J.~Andrews},
\newblock \bibinfo{title}{Optimal coupling of pv arrays to pem electrolysers in solar–hydrogen systems for remote area power supply},
\newblock \bibinfo{journal}{International Journal of Hydrogen Energy} \bibinfo{volume}{33} (\bibinfo{year}{2008}) \bibinfo{pages}{490--498}. \DOIprefix\doi{https://doi.org/10.1016/j.ijhydene.2007.10.040}.
\bibitem[{Maroufmashat et~al.(2014)Maroufmashat, Sayedin, and Khavas}]{MAROUFMASHAT201418743}
\bibinfo{author}{A.~Maroufmashat}, \bibinfo{author}{F.~Sayedin}, \bibinfo{author}{S.~S. Khavas},
\newblock \bibinfo{title}{An imperialist competitive algorithm approach for multi-objective optimization of direct coupling photovoltaic-electrolyzer systems},
\newblock \bibinfo{journal}{International Journal of Hydrogen Energy} \bibinfo{volume}{39} (\bibinfo{year}{2014}) \bibinfo{pages}{18743--18757}. \DOIprefix\doi{https://doi.org/10.1016/j.ijhydene.2014.08.125}.
\bibitem[{Godula-Jopek(2015)}]{godula2015hydrogen}
\bibinfo{author}{A.~Godula-Jopek}, \bibinfo{title}{Hydrogen production: by electrolysis}, \bibinfo{publisher}{John Wiley \& Sons}, \bibinfo{year}{2015}.

\end{thebibliography}

\end{document}